\documentclass[12pt]{article}
\usepackage{graphics}
\usepackage{amsmath}
\usepackage{amssymb}
\usepackage{graphicx}
\usepackage{makeidx}
\usepackage{mathrsfs}
\usepackage{amsfonts}
\usepackage[mathscr]{eucal}
\usepackage{amsmath}

\def\slantfrac#1#2{\hbox{$\,^#1\!/_#2$}}
\def\pni{\par\noindent}
\def\vsh{\smallskip}

\def\vsp{\vsh\pni} 

\def\cen{\centerline}
    
\def\q{\quad} 
\def \rec#1{\frac{1}{#1}}
\def\ds{\displaystyle}
\def\eg{{\it e.g.}\ }
\def\ie{{\it i.e.}\ }

\def\e{{\rm e}}
 \def\E{{\rm E}}
\def\tilde{\widetilde}

\def\Js{{\widetilde J(s)}}
\def\Gs{{\widetilde G(s)}}

\def\L{{\mathcal L}} 
\def\NN{{\bf N}}
\def\RR{{\bf R}}
\def\CC{{\bf C}}

\begin{document}

\cen{{\bf FRACALMO PRE-PRINT: \ http://www.fracalmo.org}}
\cen{{\bf The European Physical Journal, Special Topics, Vol. 193  (2011) 133--160}}
\cen{{\bf Special issue:   Perspectives on Fractional Dynamics and Control}}
\cen{{\bf Guest Editors: Changpin LI  and Francesco MAINARDI }}
\vsh
\hrule

\vskip 0.50truecm
\font\title=cmbx12 scaled\magstep2
\font\bfs=cmbx12 scaled\magstep1
\font\little=cmr10
\begin{center}
{\title Creep, Relaxation and Viscosity Properties}
\\ [0.25truecm]
{\title for Basic Fractional Models in Rheology}
\\ [0.25truecm]
{Francesco MAINARDI} $^{(1)}$ and
{Giorgio SPADA}$^{(2)}$
\\
$\null^{(1)}$
 {\little Department of Physics, University of Bologna, and INFN,} \\
{\little Via Irnerio 46, I-40126 Bologna, Italy} \\
{\little Corresponding Author. E-mail: francesco.mainardi@unibo.it} 
\\ [0.25 truecm]
$\null^{(2)}$
{\little Dipartimento di Scienze di Base e Fondamenti, University of Urbino} \\
{\little  Via Santa Chiara 27, I-61029 Urbino, Italy}
\end{center}
\begin{abstract}
\noindent
%
The purpose of this paper is  twofold: from one side
we  provide a general survey 
to the viscoelastic models constructed via fractional calculus
and from the other side  we intend to analyze   the  basic fractional models
 as far as their creep, relaxation and viscosity properties are considered.
 The basic models are those that generalize via derivatives of fractional order 
 the classical mechanical models 
characterized by  two, three and four parameters, that we refer to as 
 Kelvin--Voigt, Maxwell, Zener, anti--Zener and Burgers.
 For each fractional model we  provide plots of the creep compliance, relaxation modulus 
 and effective viscosity in non dimensional form in terms of a suitable time scale
 for different values of the order of fractional derivative. 
 We also discuss the role of the order of fractional derivative in modifying 
 the properties of the classical models.
  \end{abstract}
{\it 2010 Mathematics Subject Classification (MSC)}:
26A33, 33E12,  44A10
\vsp
{\it Physical and Astronomy  Classification Subject (PACS)}:
\vsp

%
\section{Introduction}
\label{intro}
A topic of continuum mechanics, where {fractional calculus}
is suited to be applied, is without doubt
the linear theory of viscoelasticity.
In fact, an increasing  number of authors have  
used {fractional calculus} as an empirical method of describing the
properties of linear visco\-elastic  materials. 
A wide bibliography up to nowadays is contained in the recent book 
by Mainardi \cite{Mainardi_BOOK10} including an historical perspective up to 1980's.
\vsp
The purpose of this paper is to provide, after a general survey  
to the linear theory of viscoelasticity,
a (more) systematic discussion and a graphical representation of the main properties
of the basic models described by stress--strain relationships of fractional order.
The properties under discussion  concern the standard creep and  relaxation tests that have
a relevance in experiments.
\vsp     
The plan of paper is as following. 
In  Section 2 we recall the  essential notions of linear viscoelasticity
in order to present our notations for the analog mechanical models.
We limit our attention to the basic mechanical models,
characterized by  two, three and four parameters, that we refer to as 
 Kelvin--Voigt, Maxwell, Zener, anti--Zener and Burgers.
\vsp
In Section 3 we  consider  our main topic
concerning the creep, relaxation and viscosity properties  of the  previous basic models 
generalized by replacing in their
differential constitutive equations
the  derivatives of integer order $1$ and $2$  with derivatives of fractional order $\nu$ and 
$1+\nu$ respectively,  with $0<\nu\le 1$. 
We provide the analytical expressions and the plots of the creep compliance,
relaxation modulus and effective viscosity for all the considered fractional models.
\vsp
We also enclose two Appendices for providing the readers with the essential notions of fractional derivative
and with a discussion on initial conditions.

\section{Essentials  of linear viscoelasticity}

 In this section  we present the fundamentals of  linear
 viscoelasticity restricting our attention to the one--axial case
and assuming that the viscoelastic body is quiescent for all times prior to
some starting instant that we assume as $t=0$.
\vsp
For the sake of convenience both stress $\sigma(t)$ and strain $\epsilon(t)$ are intended
to be normalized, \ie scaled  with respect to a suitable reference state
$\{\sigma _0\,, \,\epsilon _0\}\,. $

\subsection{Generalities} 
According to the linear theory, the viscoelastic body
can be considered as a linear system
with the stress (or strain) as the excitation function (input)
and the strain (or stress) as the response function (output).
 In this respect, the response functions to
an excitation expressed by the Heaviside step function $\Theta(t)$
are known  to   play a fundamental role both from a mathematical
and physical point of view. We denote by $J(t)$ the strain response
to the unit step of  stress,  according to the {\it creep test}
 and by $G(t)$ the stress response to a unit step of strain,
according to the {\it relaxation test}.
\vsp
The functions $J(t)$ and $G(t)$  are usually referred to as the
{\it creep compliance} and {\it relaxation modulus}
respectively, or, simply, the {\it material functions}
of the viscoelastic body.   In view of the causality
requirement, both  functions are    vanishing
for $t<0$.
\vsp
The limiting values of the material functions
for $t \to 0^+$ and $t \to +\infty$  are related to the
instantaneous (or glass) and equilibrium behaviours of the viscoelastic
body, respectively. As a consequence, it is usual to denote
   $J_g := J(0^+)$  the {\it glass compliance},
   $ J_e := J(+\infty)$ the {\it equilibrium  compliance},
and
  $ G_g := G(0^+)$ the {\it glass modulus}
  $ G_e := G(+\infty)$ the {\it equilibrium modulus}.
As a matter of fact, both the material functions are
non--negative. Furthermore, for  $0< t < +\infty\,,$
 $ J(t)$ is a  {\it non decreasing function} and
 $G(t)$ is a  {\it non increasing function}.
\vsp
The  monotonicity properties of $J(t)$ and $G(t)$
are  related respectively   to the physical phenomena
of  strain {\it creep} and stress {\it relaxation}.
 We also note that in some cases the material functions can contain terms  represented by 
{\it generalized functions} (distributions) in the sense of Gel'fand--Shilov \cite{Gelfand-Shilov_BOOK64}
or {\it pseudo--functions} in the sense of Doetsch \cite{Doetsch_BOOK74}.
\vsp
Under the hypotheses of causal histories, we get the stress--strain
relationships
$$
\left\{
\begin{array}{ll}
{\ds \epsilon (t)}
&= {\ds \int_{0^-}^t  \!\! J(t-\tau)\, d\sigma (\tau )}
= {\ds \sigma (0^+)\, J(t) + \int_0^t  \!\! J(t-\tau)\,
 \frac{d}{d\tau }\sigma (\tau ) \, d\tau }\,,
  \\
{\ds \sigma  (t)}
&= {\ds \int _{0^-}^t  \!\!G(t-\tau )\, d\epsilon  (\tau )}
=  {\ds \epsilon(0^+)\, G(t) + \int_0^t  \!\! G(t-\tau)\,
 \frac{d}{d\tau } \epsilon  (\tau ) \, d\tau}\,,
 \end{array}
 \right.
 \eqno(2.1)$$
 where the passage to the RHS is justified
 if differentiability is assumed for the stress--strain histories,
 see also the excellent book by Pipkin \cite{Pipkin_BOOK86}.
Being of convolution type,  equations (2.1) can be conveniently
treated by the technique of Laplace transforms so they read
in the Laplace domain
$$ \widetilde \epsilon (s) = s\, \widetilde{J}(s) \, \widetilde \sigma(s)\,,
\quad
\widetilde \sigma  (s) = s\, \widetilde{G}(s) \, \widetilde \epsilon (s)\,,
\eqno(2.2)$$
from which we derive the {\it reciprocity relation}
  $$ s\, \widetilde{J}(s)  = \frac{1}{s\,\widetilde{G}(s)}
 \,. \eqno(2.3)$$
Because of the limiting theorems for the Laplace transform, we deduce that
$ J_g = {1}/ {G_g}$, $\, J_e = {1}/{G_e}$,
with the convention that $0$ and $+\infty$ are reciprocal to each other.
The above remarkable relations allow us to classify the viscoelastic bodies
according to their instantaneous and equilibrium responses
in four types as  stated by Caputo and Mainardi in their 1971 review paper
 \cite{Caputo-Mainardi_71RNC}, see  Table 2.1.
 \vskip 0.25truecm
\begin{center}
\begin{tabular}{|c||c|c||c|c|}
\hline
Type & $J_g$ & $J_e$ & $G_g$ & $G_e$ \\
\hline
I   & $>0$ & $<\infty$ & $<\infty$ & $>0$ \\
II  & $>0$ & $=\infty$ & $<\infty$ & $=0$ \\
III & $=0$ & $<\infty$ & $=\infty$ & $>0$ \\
IV  & $=0$ & $=\infty$ & $=\infty$ & $=0$ \\
\hline
\end{tabular}  
\vskip 0.25truecm
 Table 2.1 \ The four types of viscoelasticity.
\end{center}
\vsp
 We note that the viscoelastic bodies of type I exhibit
both instantaneous and equilibrium elasticity, so their behaviour
appears  close to the purely elastic one for sufficiently short and
long times. The  bodies of  type II and IV exhibit a complete stress
relaxation (at constant strain) since $G_e =0$ and an infinite
strain creep (at constant stress) since $J_e = \infty\,,$ so  they
do not present equilibrium elasticity. Finally, the bodies of type
III and IV do not present instantaneous elasticity  since $J_g =
0\,$ ($G_g =\infty$). Other properties will be pointed out later on.
\subsection{The mechanical models} 
To get some feeling for linear
viscoelastic behaviour, it is useful to consider the simpler
behaviour of analog {\it mechanical models}. They are
 constructed from linear springs and dashpots,
disposed singly and in branches of two (in series or in parallel).
As analog of stress and strain, we use the total extending force
and the total extension.
We note that when two  elements are combined in series [in parallel],
their compliances [moduli] are additive. This can be stated
as  a combination rule: {\it creep compliances add in  series,
while relaxation moduli add in parallel}.
\vsp
 The mechanical models play an important role in the literature which is
 justified by the historical development.
 In fact, the early theories were established with the aid of these
 models, which are still helpful to visualize properties and laws of the
 general theory, using the combination rule.
\vsp
Now, it is worthwhile to consider the simple models of Figs. 1, 2, 3
  providing  their governing  stress--strain
relations along with the related material functions.
 \paragraph{The Hooke and the Newton models.}
The spring (Fig. 1a) is the elastic (or storage) element, as for it the
force is proportional to the  extension;
it represents a perfect elastic body obeying the Hooke
law (ideal solid).
This model is thus referred to as  the {\it Hooke} model.
If we denote by $m$ the pertinent elastic modulus  we have
$$
 Hooke \; model\;: \q \sigma(t)  = m\, \epsilon (t)\,, \quad\hbox{and}\quad
\left\{ 
\begin{array}{ll}
 J(t) &= 1/m \,,\\
 G(t) &= m  \,.
 \end{array}
 \right .
 \eqno(2.4)$$
In this case we have no creep and no relaxation
so the creep compliance and the relaxation modulus are constant functions:
$J(t) \equiv J_g \equiv J_e =1/m$; $G(t)\equiv G_g \equiv G_e = m$.
\vsp
The dashpot (Fig. 1b) is the viscous (or dissipative) element,
the force being proportional to rate of extension;
it represents a perfectly viscous body obeying the Newton law (perfect
liquid).
This model is thus referred to as  the {\it Newton} model.
If we denote by $b_1$ the pertinent  viscosity coefficient, we have
    $$  Newton \; model\;: \q
\sigma(t)  = b_1\, \frac{d\epsilon}{dt}\,, \quad\hbox{and}\quad
  \left\{  
  \begin{array}{ll}
J(t) &={\ds {t}/{b_1}}\,,\\
 G(t) &= b_1 \, \delta (t)\,.
\end{array}\right .
 \eqno(2.5)$$ 
 In this case we have a linear creep $J(t)= J_+t$ and instantaneous relaxation
 $G(t)= G_-\, \delta(t)$ with $G_-= 1/J_+ = b_1$.
 \vsp
We note that
 the {\it Hooke} and {\it Newton} models represent the limiting cases of
viscoelastic  bodies of type $I$ and $IV$, respectively.
 \paragraph{The Kelvin--Voigt and the Maxwell models.}
A branch constituted by a spring in parallel with a dashpot is known as
  the   {\it Kelvin--Voigt}  model (Fig. 1c). We have
$$ Kelvin-Voigt \; model \,:\;
\sigma(t)  = m\, \epsilon (t) +b_1\, \frac{d\epsilon}{dt}\,, \eqno (2.6a)
$$
and
$$\left\{ \begin{array}{ll}
 {\ds J(t)} = {\ds J_1 \left[ 1-\e^{\ds - t/\tau _\epsilon}\right]}\,, &
  {\ds J_1 = \frac{1}{m}\,,\; \tau _\epsilon  = \frac{b_1}{m}}\,,\\ \\
 {\ds G(t) = G_e +  G_- \,  \delta(t)} \,,&
   {\ds G_e=m\,,\; G_- =b_1}\,,
   \end{array} \right .\eqno(2.6b)
 $$
where $\tau _\epsilon$ is referred to as the {\it  retardation time}.
\vsp
A branch constituted by a spring in series with a dashpot is known as
 the  {\it  Maxwell} model  (Fig. 1d). We have
$$
 Maxwell \; model\,: \;
\sigma(t) +a_1 \, \frac{d\sigma}{dt}    = b_1\, \frac{d\epsilon}{dt}\,,
\eqno(2.7a)$$
and
$$\left\{ \begin{array}{ll} 
 {\ds J(t) =  J_g + J_+\,t}\,, & {\ds J_g= \frac{a_1}{b_1} \,,\; J_+ = \frac{1}{b_1}}\,,  \\ \\
 {\ds G(t) = G_1\,\e^{\ds -t/\tau_\sigma}}\,, & {\ds G_1=\frac{b_1}{a_1}\,,\; \tau _\sigma= a_1}\,,
\end{array} \right . \eqno(2.7b)
$$
where $\tau _\sigma$ is  referred to as  the {\it relaxation time}.
\vsp
\begin{figure}[h!]
\begin{center}
 \includegraphics[width=.75\textwidth]{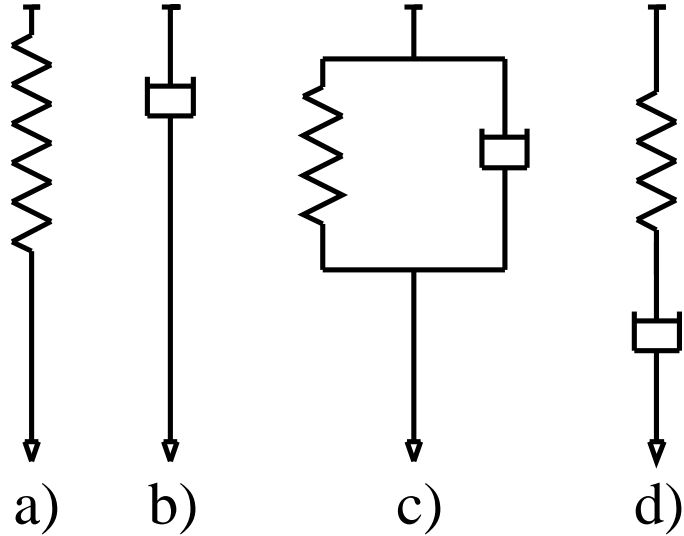}
\end{center}
\caption{The representations of the basic mechanical models:
 a)  spring for Hooke, b) dashpot for Newton,
 c)  spring and dashpot in parallel for Voigt,
 d) spring and dashpot in series for Maxwell.}
\end{figure}
\vsp
 The {\it Voigt} and the {\it Maxwell} models are thus the simplest
viscoelastic bodies of type   $III$ and $II$, respectively.
The {\it Voigt} model exhibits  an exponential  (reversible)
strain creep but no stress relaxation; it is also referred to
as the retardation element.
The {\it Maxwell} model exhibits  an exponential (reversible) stress
relaxation and a linear (non reversible) strain creep; it is
also referred to as the relaxation element.
\paragraph{The Zener and the anti--Zener models.}
Based on the combination rule,
 we can continue the previous procedure in order to construct
 the simplest models of type $I$ and $IV$ that require three parameters.
\vsp
 The simplest viscoelastic body of type $I$
 is obtained by
adding a spring either in series to a Voigt model  or
in parallel to a Maxwell model (Fig. 2a and Fig. 2b, respectively).
So doing, according to the combination
rule, we add a positive constant both to the  Voigt--like creep compliance
and to the  Maxwell--like relaxation modulus so that we obtain $J_g >0$
and $G_e >0\,. $
Such a model was
introduced by Zener \cite{Zener_BOOK48} with the denomination of {\it Standard
Linear Solid} ($S.L.S.$). We have
$$
Zener \; model \;:\q
\left[1 +a_1 \, \frac{d}{dt}\right] \sigma(t) =
 \left [ m+ b_1\, \frac{d}{dt}\right] \epsilon (t) \,,
 \eqno(2.8a) $$
and
$$
\!\!\!
 \left\{ \begin{array}{ll}
 {\ds J(t) =  J_g + J_1  \left[ 1-\e^{\ds - t/\tau_\epsilon}\right]},
      &
{\ds J_g =  \frac{a_1}{b_1}, \; J_1 =\frac{1}{m}- \frac{a_1}{b_1},\; \tau_\epsilon =\frac{b_1}{m}},
 \\
 {\ds G(t) = G_e + G_1 \,\e^{\ds -t/\tau_\sigma} },
        &
 {\ds G_e =  m, \;
   G_1 = \frac{b_1}{a_1}- m , \;  \tau_\sigma = a_1}\,.
\end{array} \right . \eqno(2.8b)
$$
We point out the condition
$ 0< m<b_1/ a_1$
in order  $J_1 ,G_1 $ be positive and hence
$0< J_g <  J_e < \infty $ and $0 <G_e <G_g < \infty\,.$
As a consequence,
we note  that, for the {\it S.L.S.} model,
 the retardation time must be greater than
the relaxation time, \ie
$\, 0 < \tau _\sigma <\tau_\epsilon < \infty \,.$
\vsp
\begin{figure}[h!]
\begin{center}
 \includegraphics[width=.82\textwidth]{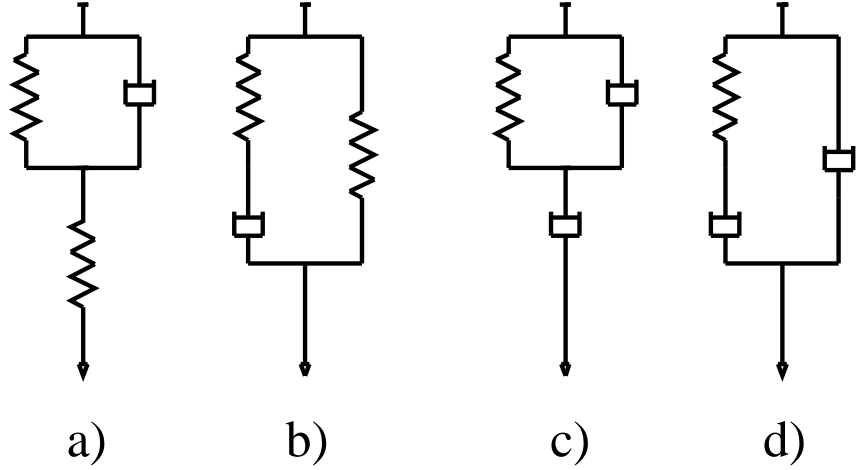}
\end{center}
\caption{The mechanical representations of the Zener model, see a), b) and
of the anti--Zener model, see  c), d), where:
a)  spring in series with  Voigt, b) spring  in parallel with Maxwell,
c)  dashpot in series with  Voigt, d) dashpot in parallel with Maxwell.}
\end{figure}
\vsp
Also the simplest viscoelastic body of type $IV$ requires three parameters,
\ie   $a_1\,,\, b_1\,,b_2\,$; it  is obtained adding a dashpot either
 in series to a Voigt model  or in parallel
 to a Maxwell model (Fig. 2c and Fig 2d, respectively).
According to the combination
rule, we add a linear term  to the  Voigt--like creep compliance
and a delta impulsive term to the  Maxwell--like relaxation modulus so that
we obtain $J_e = \infty$ and $G_g = \infty\,. $
We may refer to this model to as the
 {\it anti--Zener} model. 
  We have
$$
anti-Zener\;model\;:  \quad
\left[1 +a_1 \, \frac{d}{dt}\right] \sigma(t) =
 \left [  b_1\, \frac{d}{dt} + b_2\, \frac{d^2}{dt^2}\right] \epsilon (t)\,,
\eqno(2.9a) $$
and
$$
\!\!
 \left\{
 \begin{array}{ll}
 {\ds J(t) \!= \! J_+t + J_1 \left[ 1-\e^{\ds-t/\tau_\epsilon}\right]},
 & 
 {\ds J_+ \!=\! \frac{1}{b_1}, \,
 J_1 \!=\!\frac{a_1}{b_1}- \frac{b_2}{b_1^2},\,
       \tau_\epsilon \!=\!\frac{b_2}{b_1}}, \\
 {\ds G(t) \!= \! G_-\, \delta (t) + G_1 \,\e^{\ds -t/\tau_\sigma}},
 & 
 {\ds G_- \!=\! \frac{b_2}{a_1}, \,
   G_1 = \frac{b_1}{a_1}- \frac{b_2}{a_1^2} , \,
       \tau_\sigma = a_1}.
\end{array} \right . \eqno(2.9b)
$$
We point out the condition
$ 0< b_2/b_1< a_1$
in order  $J_1 ,G_1 $ be positive.
 As a consequence,
we note  that, for the {\it anti--Zener} model,
 the relaxation time must be greater than
the retardation time, \ie
$\, 0 < \tau_\epsilon <\tau_\sigma < \infty \,,$
on the contrary of the Zener ($S.L.S.$) model.
 \vsp
In Fig. 2 we exhibit the mechanical representations of the Zener model (2.8a)-(2.8b),
see a), b), and of the anti--Zener model (2.9a)-(2.9b), see c), d).
\vsp
 {\bf Remark:}
 We note that the constitutive equation of the anti--Zener model is formally obtained from that of the Zener model 
 by replacing the strain $\epsilon(t)$ by the stain--rate $\dot \epsilon(t)$.
 However  the Zener model, introduced by Zener in 1948 \cite{Zener_BOOK48} for anelastic metals,  
 was formerly introduced  by  Jeffreys with respect to bodily imperfection of elasticity in tidal friction, 
 since from the first 1924 edition of his treatise on the Earth \cite{Jeffreys_EARTH24,Jeffreys_EARTH70},
 Sir Harold Jeffreys was usual to refer to the rheology of the Kelvin--Voigt model 
 (that was suggested to him by Sir J. Larmor)
 to as {\it firmoviscosity} and to the rheology of the Maxwell model to as {\it elastoviscosity}.  
 We observe that  in the literature of rheology of viscoelastic fluids (including polymeric liquids) 
 our anti--Zener  model is (surprisingly for us)  known  as {\it Jeffreys fluid}, 
 see e.g. the review paper by Bird and Wiest \cite{Bird-Wiest_95}.  
 Presumably this is  due to the replacement of the strain with the strain--rate
 (suitable for fluids) in the stress--strain relationship introduced by Jeffreys.  
 As a matter of fact, in Earth rheology the Jeffreys model is known to be  the creep model
  introduced by him in 1958, see \cite{Jeffreys_58}, as  generalization of the 
  Lomnitz logarithmic creep law and well described in the subsequent  editions
 of Jeffreys' treatise on the Earth.  In view of above considerations, we are tempted to call our 
 anti--Zener model as {\it Standard Linear Fluid} in analogy with  the terminology
 {\it Standar Linear Solid} commonly  adopted for the Zener model.    
\paragraph{The Burgers model.}
In Rheology literature it is customary to consider the so--called
{\it Burgers model},
which is  obtained
by adding a  dashpot or a spring to the 
representations of the Zener or of the anti--Zener model, respectively.
Assuming   the creep representation the dashpot or the spring 
is added in series, so the Burgers model  results in 
a series combination of a Maxwell element with a Voigt element. 
Assuming the   relaxation representation,
the dashpot or the spring is added in parallel,
so the Burgers model results  in two Maxwell elements disposed in parallel.
We refer the reader to  Fig. 3 for the two mechanical representations
of the Burgers model.
\vsp
 According to our general classification,
  the Burgers model is thus a four--element  model 
of type II, defined by the four parameters $\{a_1, a_2, b_1,b_2\}$.
\vsp
 We have
$$ \!\! Burgers \; model:  \,
\left[1 +a_1  \frac{d}{dt} +a_2\frac{d^2}{dt^2} \right] \sigma(t) \! =\!
  \left [b_1 \frac{d}{dt} + b_2 \frac{d^2}{dt^2}\right] \epsilon (t),
\eqno(2.10a) $$
so 
$$
 \left\{ \begin{array}{ll}
 {\ds J(t)} &= {\ds  J_g + J_+\,t + J_1 \left(1-\e^{\ds-t/\tau_{\epsilon}}\right)}\,, 
\\                   
 {\ds G(t)} &= {\ds   G_1 \,\e^{\ds -t/\tau_{\sigma, 1}} +G_2 \,\e^{\ds -t/\tau_{\sigma, 2}} }
 \,.
\end{array} \right.
    \eqno(2.10b)$$
\vsp
We leave to the reader  to express as an exercise the physical quantities 
$J_g$, $J_+$, $\tau_{\epsilon} $ and  $G_1$, $\tau_{\sigma, 1}$, $G_2$, $\tau_{\sigma,2}$,
in terms of the four parameters $\{a_1, a_2, b_1,b_2\}$ in the operator equation (2.10a). 
\begin{figure}[h!]
\begin{center}
\includegraphics[width=.40\textwidth]{{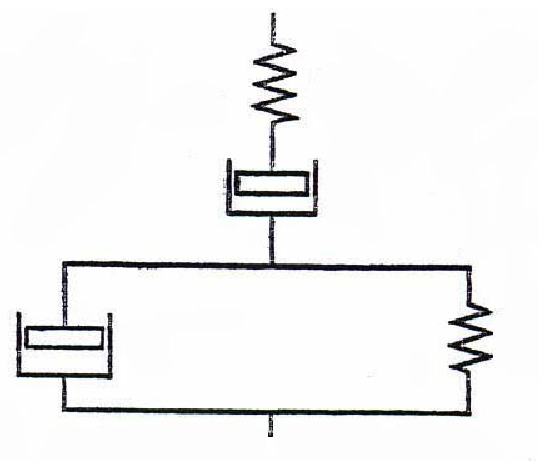}}
\vskip -0.2truecm
\includegraphics[width=.40\textwidth]{{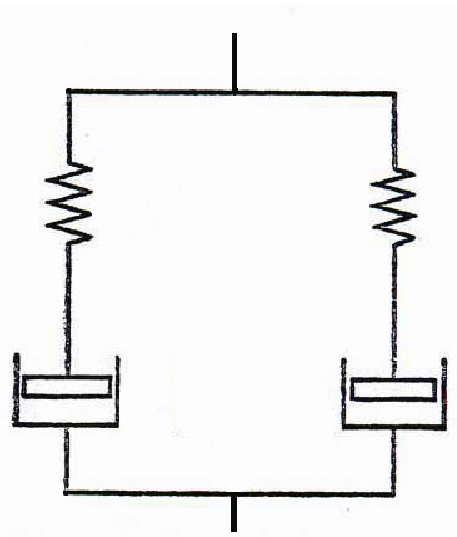}}
\end{center}
\vskip -0.2truecm
\caption{The mechanical representations of the Burgers model:
  the creep representation (top) and
 the relaxation representation (bottom).
\label{fig.2.5}}
\end{figure} 
\paragraph{The operator equation for the mechanical models}
Based on the combination rule, we can construct
models whose  material functions    are of the following type
$$\left\{ \begin{array}{ll}
   {\ds J(t)} &= {\ds J_g +\sum_{n} J_n \left[
 1-\e^{\,{\ds -t/\tau_{\epsilon,n}}}\right]
        + J_+\, t }\,,\\
  {\ds G(t)}  &= {\ds G_e +  \sum_{n} G_n \,\e^{\,{\ds - t/\tau_{\sigma,n}}}
        + G_-\, \delta (t)}\,,
		\end{array}
		\right .
         \eqno(2.11) $$
where all the coefficient are non--negative, and
 interrelated because of the
{\it reciprocity relation} (2.3) in the Laplace domain.
We note that the four types of viscoelasticity  of Table 2.1
are obtained from Eqs. (2.11) by taking into account that
$$ \left\{ \begin{array}{ll}
  J_e <\infty & \iff \;J_+ =0\,, \q \;J_e =\infty \iff J_+ \ne 0\,, \\
  G_g <\infty &   \iff G_- =0\,, \q  G_g =\infty \iff G_- \ne 0\,.
  \end{array} \right .
\eqno(2.12)
$$
Appealing to the theory of  Laplace transforms, we  write
$$
\left\{
\begin{array}{ll}
 s\Js &= {\ds J_g + \sum_n \frac{J_n}{ 1+s\,\tau_{\epsilon ,n}}
    + \frac{J_+ }{s} \,,}  \\
 s\Gs &= {\ds (G_e  + \beta) -
 \sum_n \frac{G_n}{1+s\,\tau_{\sigma,n}}
    + G_- \,s} \,,
\end{array}
\right.
 \eqno(2.13)$$
where we have put $\beta = \sum_n G_n \,. $
\vsp
Furthermore, as a consequence of (2.13),
 $\,s\Js$ and $s\Gs\,$ turn out to be {\it rational} functions
 in $\CC$ with simple poles and zeros on the negative real axis and,
possibly,
with a simple pole or with a simple zero at $s=0\,, $ respectively.
\vsp 
In these cases the integral constitutive equations (2.1) can be
written in differential form. Following Bland \cite{Bland_BOOK60} with
our notations, we obtain for these models
$$ \left[ 1+ \sum_{k=1}^p \,a_k\,{d^k\over dt ^k}\right] \, \sigma (t) =
\left[ m+ \sum_{k=1}^q \,b_k\,{d^k\over dt ^k}\right] \, \epsilon (t)\,,
   \eqno(2.14)$$
where $q$ and  $p$ are integers with $q=p$ or $q=p+1$
and $m, a_k,b_k$ are non--negative constants,
subjected to proper restrictions in order
 to meet the physical requirements of realizability.
The general Eq. (2.14) is referred to as the
{\it operator equation} of the mechanical models.
\vsp
In the Laplace domain,  we thus get
$$ s\Js  =      \rec{s\Gs}= \frac{P(s)}{Q(s)}\,,
   \q {\rm where} \q
   \begin{cases}
   {\ds P(s) = 1+ {\sum_{k=1}^{p}} \, a_k \,s^k\,,} \\ \\
   {\ds Q(s) = m+ {\sum_{k=1}^{q}} \, b_k \,s^k\,.}
 \end{cases}
    \eqno(2.15)$$
with $m \ge 0$ and $q=p$ or $q=p+1\,. $
The polynomials at the numerator and denominator
 turn out to be  {\it Hurwitz polynomials} (since they have no zeros
 for $\,Re \, \{s\} >0$) whose zeros
are alternating on the negative real axis ($s\le 0$). The least zero
in absolute magnitude is a zero of $Q(s)$.
The four types of viscoelasticity then correspond to   whether the least zero
is ($J_+ \ne 0$) or is not ($J_+ =0$) equal to zero and to
whether the greatest zero in absolute magnitude is a zero of $P(s)$
($J_g \ne 0$) or a zero of $Q(s)$ ($J_g =0$).
\vsp
In Table 2.2 we summarize the four cases, which are expected to occur in the
{\it operator equation} (2.14),  corresponding
to the four  types of viscoelasticity.
\begin{center}
\begin{tabular}{|c||c|c|c|c|c|c|}
\hline
Type & $Order$ & $m$ & $J_g$ & $G_e$ & $J_+$ & $G_-$  \\
\hline
I   & $q=p$   & $>0$ & $a_p/b_p$ & $m$ & $0$     & $0$     \\
II  & $q=p$   & $=0$ & $a_p/b_p$ & $0$ & $1/b_1$ & $0$     \\
III & $q=p+1$ & $>0$ & $  0    $ & $m$ & $0$     & $b_q/a_p$\\
IV  & $q=p+1$ & $=0$ & $  0    $ & $0$ & $1/b_1$ & $b_q/a_p$\\
\hline
\end{tabular}
\vskip 0.25truecm
Table 2.2: The four cases  of the operator equation.
\end{center}
We recognize that for $p=1$, Eq. (2.14) includes the operator
equations for  the classical models with two parameters: Voigt and
Maxwell, illustrated in Fig. 1, and with three parameters: Zener and
anti--Zener, illustrated in Fig. 2. In fact we recover the Voigt
model (type III) for  $m>0$ and $p=0, q=1$, the Maxwell model (type
II) for $m=0$ and $p=q=1$, the Zener model (type I) for $m>0$ and
$p=q=1$, and the anti--Zener model (type IV) for $m=0$ and $p=1,
q=2$. 
Finally, with four parameters we can construct two  models, the former with
$m=0$ and $p=q =2$, the latter with  $m>0$ and $p=1, q=2$,
referred by Bland \cite{Bland_BOOK60} to as four--element models of the first kind and 
 of the second kind, respectively. According to our convention they are of type 
 II and III, respectively.   
We have restricted our attention to the former model, the Burgers model (type II),
illustrated in Fig. 3, because it  has found numerous applications, specially 
in geo--sciences, see \eg the books by Klausner \cite{Klausner_BOOK91} 
and by Carcione \cite{Carcione_BOOK07}. 
\subsection{Complex modulus, effective  modulus and effective viscosity}
In Earth rheology and seismology it is customary to write the one dimensional stress--strain relation
in the Laplace domain in terms of a {\it complex shear modulus} $\widehat \mu(s)$
as 
$$\widetilde \sigma(s) = 2 \widehat\mu(s)\,  \widetilde \epsilon(s)\,, \eqno(2.16)$$ 
that is expected to generalize the relation   
for a perfect elastic  solid ({\it Hooke model}) 
$$ \sigma (t) = 2 \mu_0\, \epsilon(t)\,, \eqno(2.17)$$
where $\mu_0 $ denotes the {\it shear modulus}.
\vsp
Adopting this notation we   note comparing (2.16) with (2.2)
that the
functions  $\widetilde J(s)$  and  $\widetilde G(s)$ can be expressed in terms of the complex shear modulus 
$\widehat \mu(s)$ as 
$$
\widetilde J(s)  =   \frac{1}{2s\,\widehat \mu(s)}\,,\quad
\widetilde G(s)  =  \frac{2\widehat \mu(s)}{s}\,.
\eqno(2.18)$$
As a consequence we are led to introduce 
an {\it effective modulus} defined as
$$ \mu(t) := \frac{1}{2}\left[ \frac{d}{dt}\,G(t) + G_g\right]\,, \eqno(2.19)$$
Recalling that
for perfect viscous  fluid ({\it Newton model}) we have 
$$ \sigma (t) = 2 \eta_0\, \frac{d}{dt} \,\epsilon(t)\,, \eqno(2.20)$$
where $\eta_0 $ denotes the {\it viscosity coefficient},
similarly we are led to  
define, following M\"uller \cite{Muller_1986}  
an {\it effective viscosity} as 
$$ \eta(t):= \frac{1}{2 \dot{J}(t)}\,, \eqno(2.21)$$
where the dot denotes the derivative w.r.t. time $t$.
 \vsp
  We easily recognize that  for the Hooke model (2.4) we recover $\mu(t)\equiv  \mu_0 = m/2$,
  and for the Newton model (2.5) $ \eta(t)\equiv \eta_0 =  b_1/2$.    
Furthermore we recognize that for a spring (Hooke model) the corresponding  complex modulus is a constant, that is  
$$\widehat\mu _H(s) = \mu_0 \,, \eqno(2.22)$$
whereas for a dashpot (Newton model)
 $$\widehat \mu _N(s) = \eta_0 s = \mu_0  s \tau_0\,, \eqno(2.23)$$
  where  $\tau_0= \eta_0/\mu_0$ denotes  a  characteristic  time related to viscosity.
\vsp
In order to avoid possible misunderstanding, we explicitly note hat the  complex modulus 
is not the Laplace transform of the effective modulus but its Laplace transform multiplied by $s$. 
\vsp
The appropriate form of the  complex modulus $\widehat \mu(s)$ can be obtained recalling the combination rule 
for which creep compliances add in series while relaxation moduli 
add in parallel, as stated at the beginning of subsection 2.2.
 Accordingly, 
for a serial combination of two  viscoelastic models with individual complex moduli $\widehat \mu_1(s)$ and 
$\widehat \mu_2(s)$, 
we have 
$$
\frac{1}{\widehat \mu(s)} = \frac{1}{\widehat \mu_1(s)} + \frac{1}{\widehat \mu_2(s)}\,,
\eqno(2.24) $$
whereas for a combination in parallel
$${\widehat \mu(s)} = {\widehat \mu_1(s)} + {\widehat \mu_2(s)}\,.
\eqno(2.25)$$
\vsp  
We close this section with a discussion about the definition of solid--like and fluid--like behaviour
for viscoelastic materials.
The matter is  of course subjected to personal opinions. 
\vsp
Generally one may define a fluid if it can creep indefinitely under constant stress 
($J_e=\infty$), namely when it relaxes to zero under constant deformation ($G_e=0$).
According to this view, viscoelastic models of type II and IV are fluid--like whereas models of 
type I and III are solid--like.
However, in his interesting book \cite{Pipkin_BOOK86},  Pipkin calls a solid if the integral of $G(t)$
from zero to infinity diverges. This includes those cases in which the equilibrium modulus $G_e$
is not zero. It also includes cases in which $G_e=0$ but the approach to the limit is not fast enough 
for integrability, for example $G(t)\approx t^{-\alpha}$ with $0<\alpha \le 1$ as $t \to \infty.$
\section{Fractional viscoelasticity}
The straightforward way to introduce fractional derivatives in
linear viscoelasticity is to replace
in the constitutive equation (2.5) of the Newton model
 the first derivative   with a fractional
derivative of order $\nu \in (0,1)$, that, being $\epsilon(0^+)=0$,
may be intended both in the Riemann--Liouville or Caputo sense. 
We refer the reader to  Appendix A and  Appendix B for 
the essential notions of fractional calculus for causal systems and for
a discussion on initial conditions.
\vsp
Some people call the fractional model of the Newtonian dashpot (fractional dashpot) with the
suggestive name {\it pot}: we prefer to refer such model to
 as {\it Scott--Blair\/ model}, to  honour to the scientist who already in
the middle of the past century proposed such a constitutive equation
to explain a material property that is intermediate between the
elastic modulus (Hooke solid) and the coefficient of viscosity
(Newton fluid), see \eg \cite{Scott-Blair_BOOK49}. Scott--Blair was
surely a pioneer of the fractional calculus even if he did not
provide a mathematical theory accepted by mathematicians of his
time, as pointed out in \cite{Mainardi_EPJ-ST}.
 \vsp
It is known that the creep and relaxation power laws of the Scott--Blair model  
can be interpreted in terms of a continuous spectrum of retardation and relaxation times, respectively, 
see e.g. \cite{Mainardi_BOOK10}, p. 58. Starting from these continuous spectra,
 in \cite{Papoulia-et-al_RheoActa_2010} Papoulia et al. have interpreted the fractional dashpot
 by an infinite combination of Kelvin--Voigt or Maxwell elements in series or in parallel, respectively.  
We also note that  Liu and Xu \cite{Liu-Xu_2006} have computed 
the relaxation and creep functions 
for higher--order fractional rheological models involving more 
parameters than in our  analysis. 
\subsection{Fractional derivatives in mechanical models}
   The use of {\it fractional calculus} in linear viscoelasticity
leads to generalizations of the classical mechanical models:
 the basic   Newton element  is
substituted by the more general Scott--Blair element
(of order $\nu$).
In fact,   we can construct  the class of these generalized models
from Hooke and Scott--Blair elements,
disposed singly and in branches of two (in series or in parallel).
Then, extending the procedures
of the classical mechanical models (based on springs and dashpots), we will get the
{\it fractional operator equation}
(that is an operator equation with   fractional derivatives)
 in the form which
properly generalizes (2.14), \ie
$$
\left[1+\sum_{k=1}^p\,a_k\,{d^{\,\nu_k}\over dt^{\,\nu_k}}\right] \,\sigma (t) =
\left[m+\sum_{k=1}^q\,b_k\,{d^{\,\nu_k}\over dt^{\,\nu_k}}\right] \,\epsilon (t)
 \,, \q \nu _k = k + \nu -1\,.   \eqno(3.1)$$
 so, as a generalization of (2.11),
$$ 
\left\{
\begin{array}{ll}
 {\ds  J(t)} \,= &
  {\ds J_g+\sum_{n} J_n\left\{1-\E_\nu \left[-(t/\tau_{\epsilon,n})^\nu\right]\right\}
    + J_+\,\frac{\;t^{\nu}} {\Gamma(1+\nu)} }  \,, \\
  {\ds G(t)} \,= &
  {\ds G_e +\sum_{n} G_n \,\E_\nu\left[-(t/\tau_{\sigma,n})^\nu\right]
    + G_-\,\frac{\;t^{-\nu}}{ \Gamma(1-\nu)} }\,,
   \end{array} \right.
   \eqno(3.2) $$
where all the coefficient are non--negative. Here $\Gamma$ denotes the well known Gamma function,
and $E_\nu$ denotes the Mittag--Leffler function of order $\nu$
discussed hereafter along with its generalized form in two parameters $E_{\nu,\mu}$.  
Of course, for the fractional operator equation (3.1)  the four cases
summarized in Table 2.2 are expected to occur in analogy
with the operator equation (2.14).
The definitions in the complex plane of the Mittag--Leffler functions in one and two parameters 
are provided by their Taylor powers series around $z=0$, that is  
$$ E_\nu(z):=\sum_{n=0}^\infty \frac{z^{n}} {\Gamma (\nu n+1)}
\,,  \q  E_{\nu,\mu}(z): \sum_{n=0}^\infty \frac{z^{n}} {\Gamma (\nu n+\mu)}
\,, \q \nu\,,\mu >0\,, \eqno(3.3)$$
related between them by the following expressions, see e.g. \cite{Mainardi_BOOK10}, Appendix E,
$$  E_\nu (z) = E_{\nu,1} (z) = 1 + z \,E_{\nu,1 +\nu}  ( z)\,,
 \; \frac{d}{dz}\, E_{\nu}  (z^{\nu}) = z^{\nu-1}\, E_{\nu,\nu }  (z^{\nu})\,.   \eqno(3.4)
$$
\vsp 
In our case the Mittag--Leffler functions appearing   in (3.2) are all of order $\nu\in (0.1]$ 
and argument real and negative, namely, they are of the type
$$   E_\nu[-(t/\tau)^\nu] =
 \sum_{n=0}^\infty (-1)^n\,\frac{(t/\tau)^{\nu n}} {\Gamma (\nu n+1)}
\,, \q 0<\nu<1\,,\q \tau>0\,, \eqno(3.5)$$
 that for $\nu=1$ reduce to $\exp (-t/\tau)$. 
\vsp
 Let us now outline some noteworthy properties of the Mittag--Leffler function (3.5) assuming for brevity $\tau=1$.
Since the asymptotic behaviours for small and large times are as following 
 $$  \E_\nu  (-t^\nu ) \sim 
 \left\{ \begin{array}{ll}
    {\ds 1- \frac{t^\nu }{ \Gamma(1+\nu)}\,,}\;& t \to 0^+\,,\\
{\ds \frac{t^{-\nu }} {\Gamma(1-\nu)} \,,}\; & t \to +\infty\,,
\end{array}
\right.
\eqno (3.6)$$
we recognize that for $t\ge 0$ the Mittag--Leffler function decays for short times 
 like a stretched exponential and for large times  with a  negative power law.
Furthermore, it turns out to be {\it completely monotonic} in $0<t< \infty$ 
(that is its derivatives of successive order exhibit alternating signs  like $\e^{-t}$),
 so it can    be expressed in term of a continuous distribution of elementary relaxation processes:
$$ E_\nu(-t^\nu)= 
\int_0^\infty \!\!\e^{-rt}\, K_\nu(r) \, dr \,,\;
 K_\nu (r) 
 = 	\rec{\pi\,r}\,
   \frac{  \sin (\nu \pi)}
    {r^{\nu} + 2 \cos  (\nu \pi) + r^{-\nu}} \ge 0
	\,. \eqno(3.7)$$
For completeness  we  also recall its relations with the corresponding Mittag--Leffler 
function of two parameters as obtained from (3.4): 
$$  E_\nu (-t^\nu) = E_{\nu,1} (-t^\nu)
     = 1 - t^\nu \,E_{\nu,1 +\nu}  ( -t^\nu)\,,
 \; \frac{d}{dt}\, E_{\nu}  (-t^{\nu}) =
- t^{\nu-1}\, E_{\nu,\nu}  (-t^{\nu})\,.   \eqno(3.8)
$$
\vsp
We recognize that in fractional viscoelasticity governed by the operator equation (3.1)
the corresponding  material functions are obtained
by using the combination rule valid for the classical mechanical models.
Their determination
is made easy if we take into account the following
{\it correspondence  principle} between the classical and fractional
mechanical models, outlined in 1971 by Caputo and Mainardi \cite{Caputo-Mainardi_71RNC}.
Taking  $ 0<\nu \le 1$ and denoting by $\tau>0$ a characteristic time related to viscosity,
such correspondence principle can be formally stated by the
following three equations where the Laplace transform pairs
are outlined as well:
$$
{\ds \delta (t/\tau) \,\div\, \tau
\;\Rightarrow \;
\frac{(t/\tau)^{-\nu}} { \Gamma (1-\nu)}\,\div \,\frac{1}{s}\, (s\tau)^\nu}\,,
\eqno(3.9)$$
$$ {\ds  t/\tau    \,\div\, \frac{1}{s} \, \frac{1}{(s\tau)}
 \;\Rightarrow \;
 \frac{(t/\tau)^\nu} {\Gamma (1+\nu )}
   \div \frac{1}{s}\, \frac{1}{(s\tau)^{\nu}}   } \,, \eqno(3.10)$$
 $${\ds \e^{\ds -t/\tau} \,\div\, \frac{\tau}{1 +s\tau}
  \;\Rightarrow \;
  \E_\nu [-(t/\tau)^\nu] \div \frac{1}{s}\,  \frac{(s\tau)^{\nu}}{1+ (s\tau)^\nu} }\,.
    \eqno(3.11)$$
\vsp
In the following, we will provide 
the creep compliance $J(t)$, the relaxation modulus $G(t)$ and the effective viscosity
$\eta(t)$
for a set of  fractional  models which properly generalize with fractional derivatives 
the basic mechanical models discussed in Section 2, that is  
 Kelvin--Voigt, Maxwell, Zener, anti--Zener, and Burgers,
by using their connections with Hooke and   Scott--Blair elements. 
Henceforth, for brevity we refer to all the basic models with their first letters, that is
$H$, $SB$, $KV$, $M$, $Z$, $AZ$ and $B$.  Our analysis will  be carried out by using the Laplace transform 
and the complex shear modulus of the elementary Hooke and Scott--Blair models. 
For this purpose we recall the constitutive equations for the $H$ and $SB$ models 
$$
 Hooke \; model\;: \; \sigma(t)  = m\, \epsilon (t) \,,
 \q Scott-Blair \; model \;: \; \sigma(t)  = b_1\, \frac{d^\nu}{dt^\nu}\epsilon (t)\,, \eqno(3.12)$$
 and, setting 
 $$ \mu := \frac{m}{2}\,,\q  \tau^\nu:= \frac{b_1}{2\mu}\,, \eqno(3.13)$$   
 we get the complex shear moduli for these elements
  $$ \widehat \mu_{H}(s) = \mu\,, \q  
\widehat \mu_{SB}(s) = \mu (s\tau)^\nu\,, \eqno(3.14)$$
where $\nu \in (0,1]$ and $\tau>0$ is a characteristic time of the $SB$ element.
So, whereas the $H$ element is characterized by a unique parameter, its elastic modulus $\mu$,  
the $SB$ element is characterized by a triplet of parameters, that is
$\{\mu, \tau, \nu \}$.  
\subsection{Fractional Kelvin--Voigt  model}\label{sec:Kelvin-Voigt}
\begin{figure}[h!]
\begin{center}
\resizebox{0.60\columnwidth}{!}{%
\includegraphics[angle=0,scale=0.60]{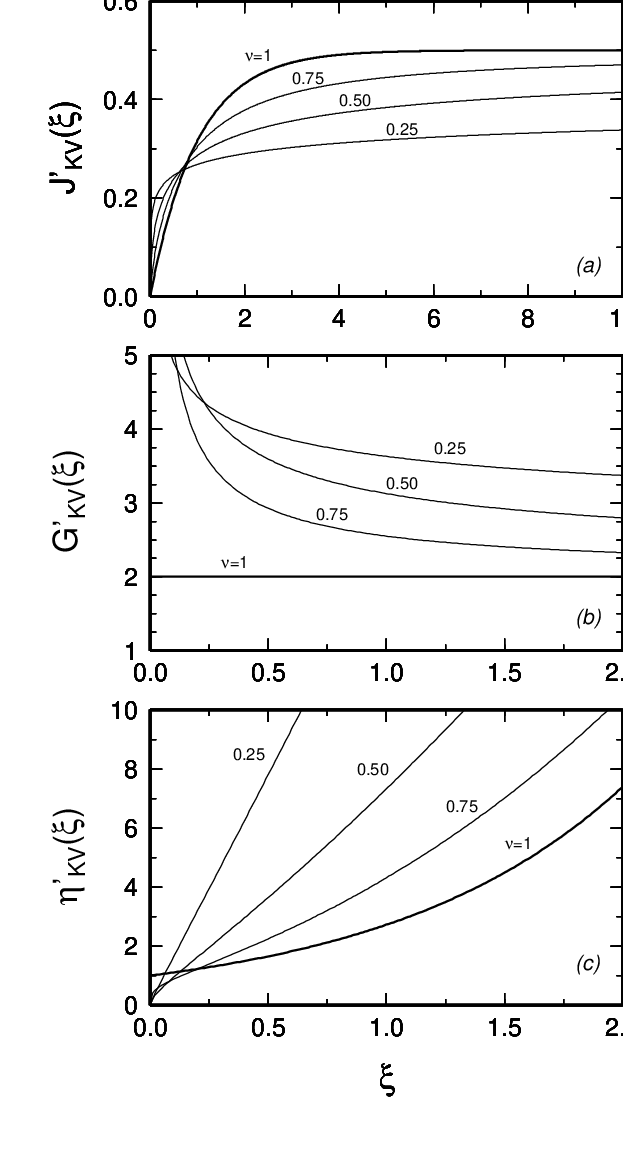} }
\end{center}
\vspace{-1cm}
\caption{Normalized creep compliance (a), relaxation modulus (b) and effective viscosity (c) for the KV body, 
for some values of the fractional power $\nu$ in the range ($0 < \nu \le 1$), as a function
of normalized time $\xi$. The thick lines ($\nu=1$) represent material functions and effective viscosity for the 
traditional Kelvin--Voingt body.}
\label{fig:Kelvin-Voigt}       
\end{figure}
\vsp
The constitutive equation for the fractional Kelvin--Voigt  model (referred to as $KV$ body) 
is obtained from (2.6a) in the form  
$$ {fractional \; Kelvin-Voigt \; model}\, : \;
\sigma(t) = m\, \epsilon (t) +b_1\, {d^\nu \epsilon\over dt^\nu }\,. \eqno(3.15)
$$
The mechanical analogue of the $KV$  body is represented 
by a Hooke ($H$) element in parallel with a Scott--Blair ($SB$) element. 
The parallel combination rule (2.25)  provides the complex modulus 
$$  
\widehat \mu_{KV}(s) = \mu \left[ 1 + (s\tau)^\nu\right]\,,
\q  \mu := \frac{m}{2}\,,\q 
\q \tau^\nu := \frac{b_1}{2\mu}\,,
\eqno(3.16)
$$  
where our time constant $\tau$ reduces for $\nu=1$ to the retardation time $\tau_\epsilon$ of the classical
$KV$ body, see (2.6b).  
Hence, substitution of (3.16) in Eqs. (2.18) gives 
$$\widetilde J_{KV}(s)  =  \frac{1}{2\mu s} 
\left[1 - \frac{(s\tau)^\nu}{1+(s\tau)^\nu}\right]\,, \q
\widetilde G_{KV}(s)  =   \frac{2\mu}{s} \left[ 1 + (s\tau)^\nu\right]\,. 
\eqno(3.17)$$
By inverting the above Laplace transforms according to Eqs. (3.11) and (3.9), we get for $t\ge 0$,
$$J_{KV}(t)  =  \frac{1}{2\mu} \left[1 - E_\nu (-(t/\tau)^\nu)\right] \,,\q
G_{KV}(t)  =  2 \mu\left[ 1 + \frac{(t/\tau)^{-\nu}}{\Gamma(1-\nu)} \right]\,. 
\eqno(3.18)$$
For plotting purposes, it is convenient to introduce 
a non--dimensional time $ \xi := t/\tau$ 
and define normalized, non--dimensional material functions.  
With $J_{KV}^\prime(\xi)=\mu J_{KV}(t)|_{t=\tau \xi}$ and $G_{KV}^\prime(\xi)=(1/\mu)G_{KV}(t)|_{t=\tau \xi}$ these 
can be written in the non--dimensional form 
$$
J_{KV}^\prime(\xi)  =  \frac{1}{2} \left[1 - E_\nu (-\xi^\nu)\right] \,, \q
G_{KV}^\prime(\xi)  =  2 \left[ 1 + \frac{\xi^{-\nu}}{\Gamma(1-\nu)} \right]\,,\q
\xi = \frac{t}{\tau}\,.  
\eqno(3.19)
$$
 Finally, 
by a straightforward application of Eq. (2.21), recalling the derivative rule in (3.8) for
the Mittag--Leffler function,
the effective viscosity of the $KV$ body
turns out to be
$$
\eta^\prime_{KV}(\xi)=  \frac{\xi^{1-\nu}}{E_{\nu,\nu}(-\xi^\nu)}\,.  
\eqno(3.20) $$
Figure~\ref{fig:Kelvin-Voigt} shows plots of of $J'_{KV}(\xi)$, $G'_{KV}(\xi)$ and $\eta^\prime_{KV}(\xi)$ as a 
function of non--dimensional time $\xi$.  
For $\nu \to 1$, the response of the fractional $KV$ body reduces to that of a classical 
$KV$ body. 
Taking into account that 
$E_1(-\xi)=\textrm{e}^{-\xi}$, 
the creep compliance reduces, in this limiting case, to $J^\prime_{KV}(\xi)=
(1 - \textrm{e}^{-\xi})/2$. 
Recalling the Dirac's delta representation $\delta(t)=t^{-1}/\Gamma(0)$, 
the relaxation modulus reduces, for $\nu \to 1$, to $G_{KV}^\prime(t)=2[1+\tau \delta(t)]=
2[1+\delta(t/\tau)]$. 
Finally, for the effective viscosity
we obtain the classical exponential law $\eta^\prime_{KV}(\xi)=\textrm{e}^\xi$. We note
 the effective viscosity exceeds the classical value since the
early stage of creep. 
\subsection{Fractional Maxwell  model}\label{sec:Maxwell}
The constitutive equation for the fractional Maxwell  model (referred to as $M$ body) 
is obtained from (2.7a) in the form  
$$ {fractional \; Maxwell \; model}\, : \;
\sigma(t) +a_1 \, \frac{d^\nu \sigma}{dt^\nu} 
   = b_1\, \frac{d^\nu \epsilon}{dt^\nu}\,. \eqno(3.21)$$
\begin{figure}[h!]
\begin{center}
\resizebox{0.60\columnwidth}{!}{%
\includegraphics[angle=0,scale=0.60]{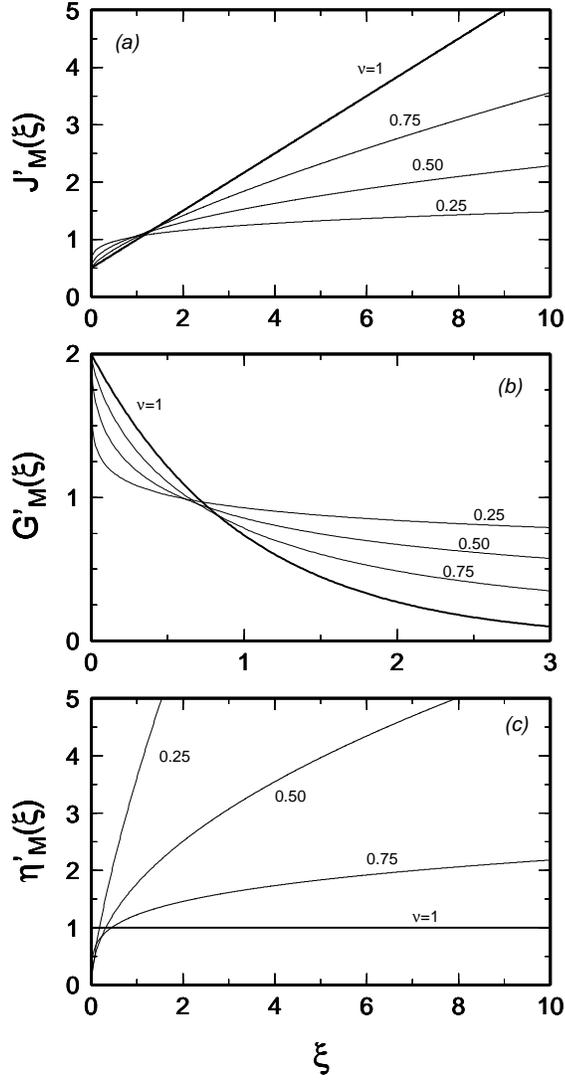} }
\end{center}
\vspace{-1cm}
\caption{Normalized creep compliance (a), relaxation modulus (b) and effective viscosity (c) for the M body, 
for some rational values of $\nu$ in the range ($0 < \nu \le 1$), as a function
of normalized time $\xi$. The thick lines, corresponding to $\nu=1$, show the 
classical Maxwell response.} 
\label{fig:Maxwell}       
\end{figure}
The mechanical analogue of the  $M$ body
is composed by a Hooke ($H$) element 
connected in series with a Scott--Blair ($SB$) element. 
 From the series combination rule (2.24) we obtain the complex modulus as 
$$\widehat \mu_M(s) = \mu \frac{(s\tau)^\nu}{1+(s\tau)^\nu}\,, \q
\mu= \frac{b_1}{2 a_1}\,, \q 
 \tau^\nu := \frac{b_1}{2\mu}\,,
\eqno(3.22)$$  
where now our time constant $\tau$ reduces for $\nu=1$ to the  relaxation time $\tau _\sigma$
of the classical $M$ body, see (2.7b).
Hence, substitution  of (3.22) into Eqs. (2.18) gives the Laplace transforms of the  material functions
$$
\widetilde J_{M}(s)  =  \frac{1}{2\mu s} \left[ 1 + \frac{1}{(s\tau)^{\nu}}\right]\,,\q
\widetilde G_{M}(s)  =   \frac{2\mu}{s} \frac{(s\tau)^\nu}{1+(s\tau)^\nu}\,.
\eqno(3.23)$$  
By inverting the Laplace transforms according to Eqs. (3.10) and (3.11), we get for $t\ge 0$, 
 $$   
J_{M}(t)  =  \frac{1}{2\mu} \left[  1 +  \frac{ \left(t/\tau\right)^\nu }{\Gamma(1+\nu)} \right]\,,
  \q
G_{M}(t)  =  2 \mu ~E_\nu \left( -\left(t/\tau\right)^\nu \right)\,.
\eqno(3.24) $$  
For plotting purposes, it is convenient to write non--dimensional forms of $J_{M}(t)$ and $G_{M}(t)$. These
can be obtained, following the example of the fractional $KV$ body, by introducing a non--dimensional time with  
$ \xi =  t/\tau$ 
and defining normalized, non--dimensional material functions  
$J_{M}^\prime(\xi)= \mu\, J_{M}(t)|_{t=\tau \xi}$ and 
$G_{M}^\prime(\xi)=(1/\mu)\, G_M(t)|_{t=\tau \xi}$, 
which provides 
$$  
J_{M}^\prime(\xi)  =  \frac{1}{2} \left[ 1 + \frac{\xi^\nu}{\Gamma(1+\nu)} \right] \, ,\q
G_{M}^\prime(\xi)  =  2 ~E_\nu \left( -\xi^\nu \right)\,, \q \xi= \frac{t}{\tau}\,. \eqno(3.25) 
$$  
Following this normalization scheme, the effective viscosity (in non-dimensional form) can be readily obtained from 
(2.21) and (3.25)  as 
$$ \eta^\prime_{M}(\xi)= \frac{\Gamma(1+\nu)}{\nu} \xi^{1-\nu}\,, \eqno(3.26)
$$  
%
Plots of $J'_M(\xi)$, $G'_M(\xi)$ and $\eta^\prime_{M}(\xi)$ are shown in 
Figure ~\ref{fig:Maxwell}. Thick lines show material functions and effective viscosity 
in the limit of $\nu \mapsto 1$, i. e., when the response of the fractional $M$ body 
degenerates into that of a classical $M$ body. In particular, 
from Eq. (3.25) we obtain 
$J^\prime_M(\xi)=(1+ \xi)/2$ and
 $G^\prime_M(\xi)=2 \textrm{e}^{-\xi}$ and the effective viscosity 
is constant ($\eta^\prime_M(\xi)=1$), which denotes the lack of transient effects. For 
$0 < \nu < 1$, the effective viscosity always increases with time and, except during the 
very early stages of creep ($\xi \approx 1$) it 
exceeds the value corresponding to the classical $M$ body ($\nu=1$). 
\subsection{Fractional Zener model}\label{sec:Zener}
\begin{figure}[h!]
\begin{center}
\resizebox{0.60\columnwidth}{!}{%
\includegraphics[angle=0,scale=0.60]{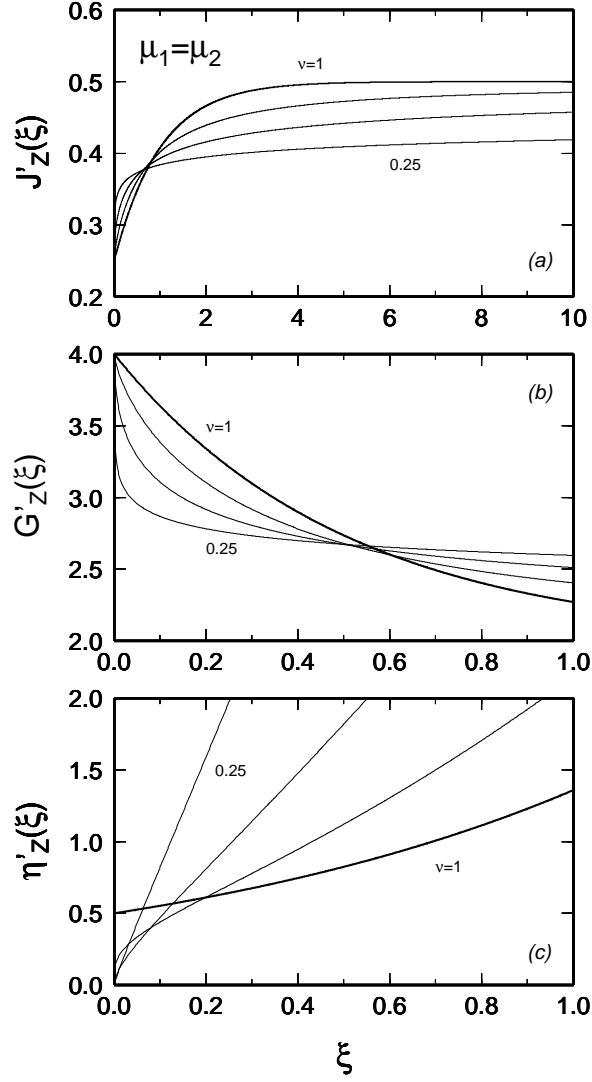} }
\end{center}
\vspace{-1cm}
\caption{Normalized creep compliance (a), relaxation modulus (b) and effective viscosity (c) for the Z body, 
for some values of the fractional power $\nu$ in the range ($0 < \nu \le 1$), as a function
of normalized time $\xi$. Here we adopt the ratio $r_\mu=\mu_2/\mu_1=1$. 
The thick lines ($\nu=1$) represent material functions and effective viscosity for the 
traditional Zener body.}
\label{fig:Zener}       
\end{figure}
The constitutive equation for the fractional Zener  model (referred to as $Z$ body) 
is obtained from Eq. (2.8a) in the form  
$$ {fractional \; Zener \; model}\, : \;
\sigma(t) +a_1 \, \frac{d^\nu \sigma}{dt^\nu} 
   = m\, \epsilon(t) +  b_1\, \frac{d^\nu \epsilon}{dt^\nu}\,. \eqno(3.27)$$
The mechanical analogue of the  fractional $Z$ body is represented by a Hooke ($H$) element in series with a 
Kelvin--Voigt ($KV$) element. 
Here we indicate with $\mu_1$ the shear modulus of the $H$ body while with 
$\{\mu_2,\tau_2,\nu\} $ the triplet of parameters characterizing   the $KV$ body. 
\vsp
 The material functions for the fractional $Z$ body 
in the time domain can be derived following the same procedure outlined above for the  fractional $KV$and $M$. 
However, the algebraic complexity increases because of the increased number of independent rheological 
parameters involved.
The use of  the combination rule (2.24), which holds for connections
in series,  provides the complex modulus 
$$
\frac{1}{\widehat \mu _{Z}(s)} = \frac{1}{\mu _1} + \frac{1}{\mu _2}\frac{1}{1 + (s\tau _2)^\nu}\,,
\q \mu _1 = \frac{b_1}{2a_1}\,, \q \mu _2 = \frac{m}{2}\,,\q  \tau _2^\nu = \frac{b_1}{2\mu _2}\,,
\eqno(3.28)$$  
where now our time constant 
$\tau_2$ reduces for $\nu=1$ to the retardation time $\tau _\epsilon$ of the classical
$Z$ body, see (2.8b).  
Hence, 
substitution of (3.28) in Eqs. (2.18)  provides the Laplace transform 
of the creep compliance and relaxation modulus. For the creep compliance
we have
$$
\widetilde J_Z(s) = \frac{1}{2s}\left[ \left( \frac{1}{\mu_1} + \frac{1}{\mu_2}\right) -\frac{1}{\mu_2}
 \frac{(s\tau_2)^{\nu}}{1+(s\tau_2)^{\nu}} \right]\,,\eqno(3.29) 
$$ 
that can be easily inverted to obtain
$$ J_Z(t) = \frac{1}{2}\left[\frac{1}{\mu_1} + \frac{1}{\mu_2}
\left( 1 - E_{\nu}(-(t/\tau_2)^{\nu}\right)\right]\,, \quad t \ge 0.  
\eqno(3.30) $$  
For the relaxation modulus the following expression can be obtained
$$\widetilde G_Z(s) = 2 \mu^* \left( \frac{\tau_2}{\tau_a} \right)^\nu \frac{1}{s} ~
\frac{s^\nu+ 1/\tau_2^\nu}{s^\nu+ 1/\tau_a^\nu}\,, 
\eqno(3.31) $$   
where 
$$\mu^\ast =  \frac{\mu_1\mu_2}{\mu_1+\mu_2}\,, 
\eqno(3.32) $$  
and we have introduced an additional   characteristic time 
$$\tau_a^\nu = \frac{1}{1+r_\mu}\tau_2^\nu\,, \q r_\mu=\frac{\mu_1}{\mu_2}\,.
\eqno (3.33)$$  
Applying a partial fraction expansion to Eq. (3.31) and using Eqs. (3.8), the relaxation
modulus in the time domain can be cast in the form
$$G_Z(t) = 2  \mu^\ast \left( \frac{\tau_2}{\tau_a} \right)^\nu 
\left[ E_\nu \left( -\left( {t}/{\tau_a} \right)^\nu \right) + 
(t/\tau_2)^\nu E_{\nu,\nu+1} \left( -\left( {t}/{\tau_a} \right)^\nu \right) \right]\,, 
\eqno(3.34)$$  
hence,  after some rearrangement, we finally obtain 
$$G_Z(t) = 2  \mu^\ast \left[ 1 + r_\mu E_\nu (-({t}/{\tau_a})^\nu)\right]. 
\eqno(3.35) $$  
We now recognize that for $\nu=1$ the constant $\tau_ a$ reduces to the relaxation time 
$\tau _\sigma $ ($0<\tau _\sigma <\tau _\epsilon <\infty$) for the classical $Z$ model, see (2.8b). 
\vsp
Writing $J_{Z}^\prime(\xi)=\mu^\ast J_{Z}(t)|_{t=\tau_2 \xi}$ and $G_{Z}^\prime(\xi)=(1/\mu^\ast)G_{Z}(t)|_{t=\tau_2 \xi}$ the
material functions for the Zener model  
can be written in the non--dimensional form 
$$
J_{Z}^\prime(\xi)  =  \frac{1}{2} \left[1 - \frac{h_1}{h_2} E_\nu (-\xi^\nu)\right]\,, 
\q G_{Z}^\prime(\xi)  =  2 \left[ 1 + h_1 E_\nu (-h_2 \xi^\nu) \right]\,, 
\q  \xi = \frac{t}{\tau_2}\,, 
\eqno(3.36)$$
with $h_1 =  r_\mu$, $h_2 =  1 + r_\mu$.
\vsp
As a consequence the effective viscosity of the $Z$ body
turns out to be
$$\eta^\prime_{Z}(\xi)=  \frac{h_1}{h_2}\frac{\xi^{1-\nu}}{E_{\nu,\nu}(-\xi^\nu)}.  
\eqno(3.37)$$ 
Figure~\ref{fig:Zener} shows plots of $J'_{Z}(\xi)$, $G'_{Z}(\xi)$ and $\eta^\prime_{Z}(\xi)$ as a 
function of $\xi$, for various values of $\nu$ in the range $0 < \nu \le 1$. In the limiting case
of $\nu=1$ (thick curves), which corresponds to the classical Zener model, the material
functions and the effective viscosity can be written in terms of exponential functions, in agreement with
Eqs. (2.8b) above. In particular,
since $r_\mu=1$, we obtain $J'_{Z}(\xi)=\slantfrac{1}{2}(1-\slantfrac{1}{2}~\textrm{e}^{-\xi})$ 
and $G'_{Z}(\xi)=2(1+\textrm{e}^{-2\xi})$.
Furthermore, the effective viscosity grows exponentially, since $\eta^\prime_{Z}(\xi)=\slantfrac{1}{2}~\textrm{e}^{\xi}$. 
We note that, for sufficiently long times ($\xi > 0.2$ in Figure~\ref{fig:Zener}), the value
of  $\eta^\prime_{Z}(\xi)$ for a fractional Zener model ($0 < \nu < 1$) always 
exceeds the classical Zener viscosity.

%
%
\subsection{Fractional anti--Zener model}\label{sec:AZ}
The constitutive equation for fractional anti--Zener  model (referred to as $AZ$ body) 
is obtained from (2.9a) in the form  
$$fractional \; anti-Zener\;model\;:  \quad
\left[1 +a_1 \, \frac{d^\nu}{dt^\nu}\right] \sigma(t) =
 \left [  b_1\, \frac{d^\nu}{dt^\nu} + b_2\, \frac{d^{1+\nu}}{dt^{1+\nu}}\right] \epsilon (t)\,.
\eqno(3.38) $$
The fractional anti--Zener model results from the combination in series of a 
$SB$ element (with material parameters $\{\mu, \tau_1, \nu$\}) 
and a fractional $KV$  element ($\{\mu, \tau_2, \nu\}$). 
\vsp
Using  the series combination rule (2.24) we obtain 
$$
\frac{1}{\widehat \mu_{AZ}(s)} \! = \!
\frac{1}{\mu} \frac{1}{(s\tau_1)^{\nu}} + \frac{1}{\mu} 
     \left[ 1 - \frac{(s\tau_2)^{\nu}}{1+(s\tau_2)^{\nu}} \right],
	 \; \frac{1}{\mu}= {2\left(\frac{a_1}{b_1} -\frac{b_2}{b_1^2}\right)},  
	 \; {\tau _1}^\nu =\frac{1}{2\mu\, b_1}, 
	 \; {\tau _2}^\nu =\frac{b_2}{b_1},  
\eqno(3.39)$$ 
where now our time constant 
$\tau_2$ reduces for $\nu=1$ to the retardation time $\tau _\epsilon$ of the classical
$AZ$ body, see (2.9b).  
Using (2.18) the above equation easily provides the Laplace transform of the creep compliance 
$$\widetilde J_{AZ}(s) =  \frac{1}{2\mu s} \left[  
\frac{1}{(s\tau_1)^{\nu}}+1  - \frac{(s\tau_2)^{\nu}}{1+(s\tau_2)^{\nu}}\right],  
\eqno(3.40) $$ 
whose inverse is
$$J_{AZ}(t) = \frac{1}{2\mu} \left[ \frac{(t/\tau_1)^{\nu}}{\Gamma(1+\nu)} 
+ \left(1- E_{\nu}(-(t/\tau_2)^{\nu})\right)\right], 
\quad t \ge 0. 
\eqno(3.41)$$
The derivation of the relaxation modulus is more complicated but straightforward. 
Starting from the general relationship (2.18) and using (3.40), 
by a partial fraction expansion we obtain  
$$\widetilde G_{AZ}(s) =  \frac{2\mu}{s} \left[ (s \tau_0)^\nu +  
\left(\frac{\tau_1}{\bar \tau}\right)^{2\nu} \frac{(s\bar\tau)^{\nu}}{1 + (s\bar\tau)^\nu}
   \right]\,,\eqno(3.42) 
$$  
where we have defined the new time constants
$$\bar \tau =(\tau_1^\nu + \tau_2^\nu)^\frac{1}{\nu} \,, \q
\tau_0 = \frac{\tau_1 \tau_2}{\bar \tau}.
\eqno(3.43)$$
Hence, using (3.9) and (3.11), the inverse Laplace transform of (3.43) is
$$G_{AZ}(t) =  2 \mu \left[ \frac{(t/\tau_0)^{-\nu}}{\Gamma(1-\nu)} + \left(\frac{\tau_1}{\bar \tau}\right)^{2\nu}
E_\nu (-(t/\bar \tau)^\nu) \right], \quad t \ge 0\,.  
\eqno(3.44)$$   
We now recognize that for $\nu=1$ the constant $\bar \tau$ reduces to the relaxation time 
$\tau _\sigma $ ($0<\tau _\epsilon <\tau _\sigma <\infty$) for the classical $AZ$ model, see (2.9b).

\begin{figure}[h!]
\begin{center}
\resizebox{0.60\columnwidth}{!}{%
\includegraphics[angle=0,scale=0.60]{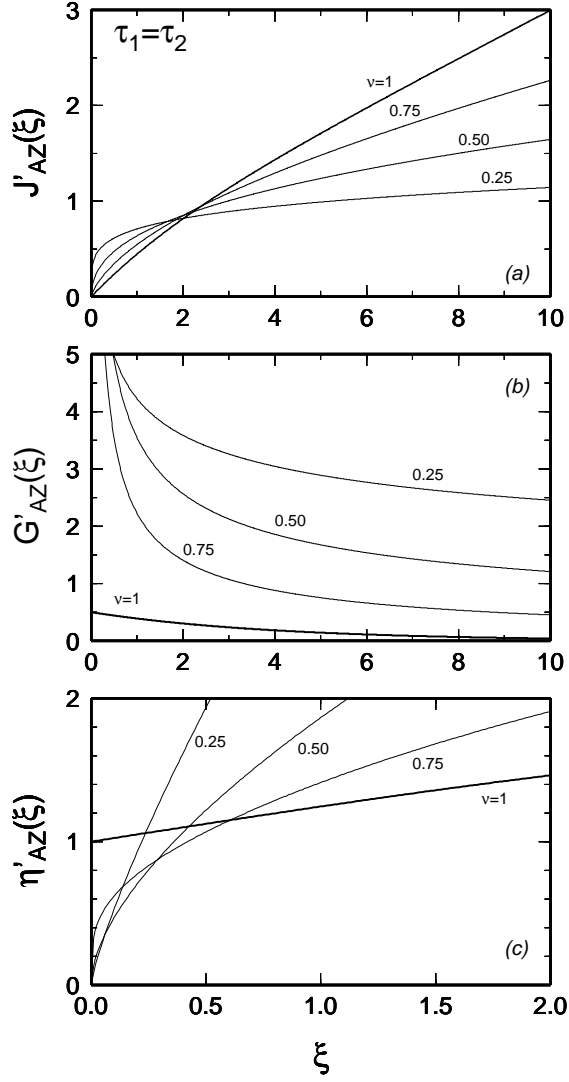} }
\end{center}
\vspace{-1cm}
\caption{Normalized creep compliance (a), relaxation modulus (b) and effective viscosity (c) for the AZ body, 
for some values of the fractional power $\nu$ in the range ($0 < \nu \le 1$), as a function
of normalized time $\xi$. 
The particular case $\tau_1=\tau_2$ is considered. The thick lines, pertaining to $\nu=1$, show results for the 
traditional AZ  body.}
\label{fig:Anti-Zener}       
\end{figure}
From (3.41) and (3.44) it is easy to derive 
non dimensional expressions for the material functions in the time domain. The choice of the
scaling for the time variable is of course arbitrary. Here we define
$$\xi =  \frac{t}{\tau^\ast} \,, \q
\tau^\ast = \frac{\tau_1 \tau_2}{\tau_1 + \tau_2} \eqno(3.45)$$
With this choice, defining $J_{AZ}^\prime(\xi)=$ $\mu J_{AZ}(t)|_{t=\tau^\ast \xi}$
and manipulating (3.41) we obtain 
$$J_{AZ}^\prime(\xi) = \frac{1}{2} \left[  
\frac{\left( c_2\xi \right)^{\nu}}{\Gamma(1+\nu)} 
+ \left(1 - E_{\nu}\left( - \left(c_1\xi \right)^{\nu} \right)\right)  \right],
\eqno(3.46)$$ 
where 
$$ c_1  =  \frac{r_\tau}{1 + r_\tau}\label{c_1} \,,\q
c_2  =  \frac{1}{1 + r_\tau} \label{c_2}\,, \q
r_\tau \equiv \frac{\tau_1}{\tau_2}\,. \eqno(3.47)$$  
Similarly, from (3.44), the relaxation modulus has 
the non dimensional form
$$
G_{AZ}^\prime(\xi) = {2} \left[ \frac{(e_1 \xi)^{-\nu}}{\Gamma(1-\nu)} +
e_2^{2\nu} E_\nu (-(e_3\xi)^\nu)\right],
\eqno(3.48) $$   
where $G_{AZ}^\prime(\xi)=$ $(1/\mu)$ $G_{AZ}(t)|_{t=\tau^\ast \xi}$ and we have
introduced the constants 
$$e_1  =  \frac{(1+r_\tau^\nu)^\frac{1}{\nu}}{1+r_\tau}\label{e1} \,,\q
e_2  =  \frac{r_\tau}{(1+r_\tau^\nu)^\frac{1}{\nu}} \label{e2}\,, \q 
e_3  =  \frac{r_\tau}{(1+r_\tau)(1+r_\tau^\nu)^\frac{1}{\nu}}\,.\label{e3}
\eqno(3.49)$$   
Using the normalization scheme above, the expression for the effective viscosity of the $AZ$ body 
can be obtained from Eqs. (3.45) and (3.46) as  
$$\eta^\prime_{AZ}(\xi)= \frac{1}{ \displaystyle{\frac{\nu}{\Gamma(1+\nu)}}c_2 \left( c_2 \xi \right)^{{\nu}-1} +
c_1 \left(c_1 \xi \right)^{{\nu}-1} E_{\nu,\nu} \left( -\left(c_1 \xi \right)^{\nu}  \right)}\,.   
\eqno(3.50)$$   
\vsp
Figure~\ref{fig:Anti-Zener} shows functions $J'_{AZ}(\xi)$, $G_{AZ}^\prime(\xi)$ and $\eta^\prime_{AZ}(\xi)$ as
a function of $\xi$, in the particular case $\tau_1=\tau_2$ ($r_\tau=1$ according to Eq. (3.46)).
The behaviour for $\nu=1$ (thick curves), which corresponds to the classical anti--Zener model, 
can be expressed in terms of elementary functions. Since $r_\tau=1$,
$c_1=c_2={1}/{2}$, see Eq. (3.47). 
Hence, from (3.46), in this particular case the normalized
creep function is $J'_{AZ}(\xi)={1}/{2}\left(1 + {\xi}/{2} + {\e}^{-\xi}\right)$, 
which,
in the range of $\xi$ values considered in Figure~\ref{fig:Anti-Zener},
 appears to be dominated by the linear term.
In the same limit, since from Eqs. (3.49) we have 
$e_1=1$, $e_2={1}/{2}$, and $e_3={1}/{4}$, omitting  
an additive $\delta(\xi)$ term, the normalized relaxation modulus decays
exponentially with $\xi$ according to
$G'_{AZ}(\xi)= \textrm{e}^{-\xi/4}/2$. 
The effective viscosity, according to (3.50), turns out to be $\eta^\prime_{AZ}(\xi)=
{2}/{(1+\textrm{e}^{-\xi/2})}$, which approximately is increasing linearly in the range
of $\xi$ values considered.  

%
%
\subsection{Fractional Burgers  model}\label{sec:B}
The constitutive equation for the fractional Burgers model (referred to as $B$ body)
is obtained from (2.10a) as
$$  fractional \; Burgers \; model:  \,
\left[1 +a_1  \frac{d^\nu}{dt^\nu} +a_2\frac{d^{1+\nu}}{dt^{1+\nu}} \right] \sigma(t) \! =\!
  \left [b_1 \frac{d^\nu}{dt^\nu} + b_2 \frac{d^{1+\nu}}{dt^{1+\nu}}\right] \epsilon (t).
\eqno(3.51) $$
The mechanical analogue of the fractional Burgers ($B$) model can be represented as the 
combination in series of a fractional $KV$ element (with material parameters $\{\mu_1, \tau_1, \nu\}$) 
and a fractional $M$ fractional element ($\{\mu_2, \tau_2, \nu\}$).
\vsp
Using the expressions (3.16) and (3.22) for complex moduli of fractional $KV$ and $M$ bodies
 into the series combination rule (2.24)
provides the complex modulus 
$$\frac{1}{\widehat \mu_{B}(s)}  
      = \frac{1}{\mu_1} \left[ 1+ \frac{1}{(s\tau_1)^{\nu}}  \right] 
     + \frac{1}{\mu_2} \left[ 1 - \frac{(s\tau_2)^{\nu}}{1+(s\tau_2)^{\nu}} \right]\,, 
	 \eqno(3.52)$$
where  the four constants $\mu_i$ and $\tau_i$ are related in some way to the four coefficients 
$a_i$, $b_i$ of the constitutive equation with $i=1,2$.   
\vsp
Using (2.18), from (3.52) it is possible to obtain immediately the Laplace transformed 
creep compliance  
$$
\widetilde J_B(s) = 
\frac{1}{2s}\left[ \left( \frac{1}{\mu_1} + \frac{1}{\mu_2}\right) + \frac{1}{\mu_1}\frac{1}{(s\tau_1)^\nu} 
-\frac{1}{\mu_2} \frac{(s\tau_2)^{\nu}}{1+(s\tau_2)^{\nu}} \right]\,, 
\eqno(3.53)$$  
which can be easily inverted with the aid of (3.10) and (3.11), obtaining
$$ J_B(t) = 
\frac{1}{2}\left[\left(\frac{1}{\mu_1} + \frac{1}{\mu_2}\right) + 
\frac{1}{\mu_1}\frac{(t/\tau_1)^{\nu}}{\Gamma(1+\nu)}  
                 - \frac{1}{\mu_2} E_{\nu}(-(t/\tau_2)^{\nu})\right]\,, \quad t \ge 0\,.  
\eqno(3.54)$$
\begin{figure}[h!]
\begin{center}
\resizebox{0.60\columnwidth}{!}{%
\includegraphics[angle=0,scale=0.60]{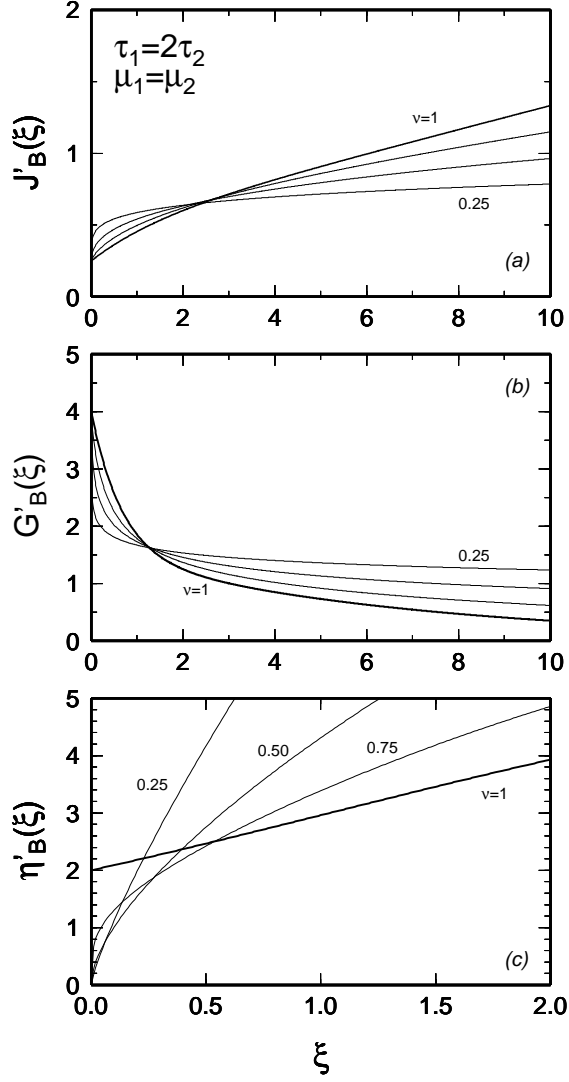} }
\end{center}
\vspace{-1cm}
\caption{Normalized creep compliance (a), relaxation modulus (b) and effective viscosity (c) for the B body, 
for values of the fractional power $\nu=1, 0.75, 0.50$ and $0.25$ as a function
of normalized time $\xi$. The thick lines ($\nu=1$), show results for the 
traditional B body.}
\label{fig:Burgers}       
\end{figure} 
\vsp
The computation of the Laplace--transformed relaxation modulus for the fractional $B$ body requires 
more cumbersome algebra compared to $J_B$. Using (3.52) into (2.18) provides,
after some algebra:
$$
\widetilde G_{B}(s) = 2 \mu^* \tau_0^\nu s^{\nu-1} \frac{z+\gamma_0}{z^2 \delta_2 + z \delta_1 + \delta_0}\,.
\eqno(3.55)$$  
Above, for the sake of convenience,
 we have introduced  the variable
$$  z\equiv (s\tau_0)^\nu\,,\eqno(3.56) $$
where $\tau_0$ is defined as in the fractional $AZ$ model (3.43), and several constants as follows:
$$ \mu^* = \frac{\mu_1 \mu_2}{\mu_1 +\mu_2}\,, \eqno(3.57)$$
$$
\gamma_0  =  \frac{r_\tau^\nu}{1+r_\tau^\nu}\,,\;
\delta_2 =  \frac{1}{1+r_\mu} \label{d2}\,,\;
\delta_1  =  \frac{1}{1+r_\mu} + \frac{r_\mu}{1+r_\mu} \cdot \frac{r_\tau^\nu}{1+r_\tau^\nu}\label{d1} \,,\; 
\delta_0  =  \frac{1}{1+r_\mu} \cdot \frac{r_\tau^\nu}{(1+r_\tau^\nu)^2} \,,\eqno(3.58)$$
where $r_\tau= \tau_1/\tau_2$ and  $r_\mu = {\mu_1}/{\mu_2}$. 
\vsp
Denoting by $z_1$ and $z_2$ the roots of algebraic equation 
$$
{z^2 \delta_2 + z \delta_1 + \delta_0} = 0\,, \eqno(3.59)$$ 
 we obtain 
$$
\tilde G_{B}(s) = 2 \mu^* \tau_0^\nu s^{\nu-1} \sum_{i=1}^2 \frac{G_i}{z-z_i}\,,
\eqno(3.60)$$
where the evaluation of constants $G_i=G_i(r_\mu,r_\tau)$ 
is straightforward (the details are left to the reader). 
Since from (3.58) it can be easily recognized that 
the discriminant of Eq. (3.59) is positive, the roots are real. Hence, observing that 
$\delta_i > 0 ~(i=0, 1, 2)$, Descartes' rule of signs ensures that $z_i < 0 ~(i=1, 2)$. Setting 
$z_i = -1/\rho_i^\nu$ with $\rho_i>0$ and recalling (3.56),  we obtain a convenient 
expression of the relaxation modulus in the Laplace domain:
$$  
\tilde G_{B}(s) = 2 \mu^* \sum_{i=1}^2 \frac{G_i}{s} \cdot \frac{(s\tau_0 \rho_i)^\nu}{1+(s\tau_0 \rho_i)^\nu}, 
\eqno(3.61)$$  
which can be easily inverted with the aid of (3.11), obtaining the 
relaxation modulus in the time domain as a combination of two Mittag--Leffler 
functions with different arguments:
$$  
G_{B}(t) = 2 \mu^* \sum_{i=1}^2 {G_i} E_\nu \left( -\left( \frac{t}{\tau_0 \rho_i} \right)^\nu \right),   
\eqno(3.62)$$   
Useful non--dimensional forms for the material functions in the time domain can be easily obtained. 
For the creep compliance (3.54) we have 
$$
J_B^\prime(\xi) = \frac{1}{2} \left[1 + c'_2 \frac{\left( c_2 \xi  \right)^{\nu}}{\Gamma(1+\nu)} - 
c'_1 E_{\nu}\left( - \left(c_1 \xi \right)^{\nu} \right)   \right]\,, 
\q \xi = \frac{t}{\tau^\ast}\,, \q \tau^* = \frac{\tau_1\, \tau_2}{\tau_1 +\tau_2}\,,
\eqno(3.63)$$
where  $J_{B}^\prime(\xi)=$ $\mu^\ast J_{B}(t)|_{t=\tau^\ast \xi}$, 
and
$$
c'_2  =  \frac{1}{1+r_\mu}\,, \q   
c'_1  =  \frac{r_\mu}{1+r_\mu}\,,\q  r_\mu = \frac{\mu_1}{\mu_2}\,, 
\eqno(3.64)$$
 For the non--dimensional form for the relaxation modulus we obtain:
$$
 G'_{B}(\xi) = 2 \sum_{i=1}^2 {G_i} E_\nu \left( - \left( \frac{e_1}{\rho_i} \xi   \right)^\nu  \right),
\eqno(3.65)$$
where $G_{B}^\prime(\xi)=$ $(1/\mu^*)$ $G_B(t)|_{t=\tau^\ast \xi}$, $\xi=t/\tau^\ast$. 
\vsp
  By the above normalization scheme, it is possible to find the expression for the effective viscosity 
of the $B$ body by using  Eqs. (2.21) and (3.63) as  
$$
\eta^\prime_{B}(\xi)= \frac{1}{ \displaystyle{\frac{\nu}{\Gamma(1+\nu)}}c_2 c'_2\left( c_2 \xi \right)^{{\nu}-1} +
	c_1 c'_1 \left(c_1 \xi \right)^{{\nu}-1} E_{\nu,\nu} \left( -\left(c_1 \xi \right)^{\nu}  \right)}.   
\eqno(3.66) 
$$
As an illustration of the behaviour of the material functions for the $B$ body, 
in Figure~\ref{fig:Burgers} we consider, for $\tau_1=2\tau_2$ and $\mu_1=\mu_2$, 
$J'_{B}(\xi)$, $G_{B}^\prime(\xi)$ and $\eta^\prime_{B}(\xi)$ as
a function of variable $\xi$. 
While the results for the creep compliance (a) and the effective viscosity (c)
 are qualitatively similar to those obtained for the AZ model in Figure~\ref{fig:Anti-Zener}, those pertaining
to the relaxation modulus (c) differ significantly, being now removed the singularity for $\xi \mapsto 0$ 
displayed in  Figure~\ref{fig:Anti-Zener}c for $\nu \ne 1$. Curves obtained for $\nu=1$ (thick), 
which corresponds to the classical B model, can be expressed in terms of elementary functions. Since $r_\tau=2$ and
$r_\mu=1$,  in this case we have $c_1={2}/{3}$, $c_2={1}/{3}$, 
$c'_1=c'_2={1}/{2}$. Hence, with these numerical values for the 
rheological parameters the creep function varies with time as 
$$J'_{AZ}(\xi)= \frac{1}{2}\, \left(
1 + \frac{\xi}{6} - \frac{1}{2}~\textrm{e}^{\ds -{2\xi}/{3}}\right)\,.\eqno(3.67)$$ 
In the same limit, since $E_1(-x)=\textrm{e}^{-x}$, the relaxation modulus 
$G'_{B}(\xi)$ reduces to the sum of two exponentially decaying functions, and its initial 
value is only determined by the numerical
value of the elastic parameters $\mu_1$ and $\mu_2$. 
The effective viscosity, according to (3.66), for $\nu=1$ asymptotically approaches a 
constant value for $\xi \to \infty$ as in the case of the  $AZ$ body; 
for all the other values of the fractional order $\eta^\prime_{B}(\xi)$
increases indefinitely, as in the case of the $M$ body. 

\section{Conclusions}

Starting from the mechanical analog of the basic fractional models of linear viscoelasticity 
 we have verified the consistency  of the {\it correspondence principle}. 
  This principle  allows one to formally get
the material functions and the effective viscosity of the fractional  models 
starting form the known expressions of the corresponding classical models by
using the corresponding rules (3.9), (3.10), (3.11). 
\vsp
In particular, in the present paper, we were able to plot all these functions versus 
a suitable time scale in order to 
visually show  the effect of the order $\nu\in (0,1]$ entering in our basic fractional models.
This can help the researchers to guess which fractional  model can better fit the experimental results.        
Presentation of the results has greatly benefited 
from a recently published Fortran code for computing the Mittag--Leffler function of complex
argument \cite{Verotta_2010a,Verotta_2010b}, coupled with an open source program for manipulating and
visualizing data sets \cite{Wessel-Smith_1998}. 
\vsp
In Earth rheology, the concept of effective viscosity is often introduced to describe the behaviour of
composite materials that exhibit both linear and a non--linear 
(i.e., non--Newtonian) stress--strain components of deformation), see
e.g., Giunchi and Spada \cite{Giunchi-Spada_2000}). 
However, according to M\"uller \cite{Muller_1986} and to the results outlined in this paper, 
it is clear that this concept has a role also within the context of linear viscoelasticity, 
both for describing the
transient creep of classical models and for characterizing the mechanical behavior of fractional models.    
\vsp
Transient rheological effects are currently the subject of investigation in the field of post--seismic 
deformations \cite{Cannelli-et-al_2010,Melini-et-al_2008,Melini-et-al_2010}, global isostatic
 deformations \cite{Spada_2008,Spada-Boschi_2006} and regional sea level variations
 \cite{Spada-Colleoni-Ruggieri_2010}. 
 The fractional mechanical models described
in this paper have the potential of better characterizing the time--dependence of these processes, also
in view of the increased quality of the available geophysical and geodetic observations.

\section{Acknowledgments}
We thank Davide Verotta for kindly providing a Fortran 90 
function (\texttt{mlfv.f90}) for the evaluation of the two--parameters 
Mittag--Leffler function $E_{\alpha,\beta}(z)$,
which has been essential for the preparation of this manuscript.  
The figures have been drawn using the Generic Mapping Tools (GMT) public domain 
software \cite{Wessel-Smith_1998}. 
This work was supported by funding  the ice2sea project from 
the European Union 7th Framework Programme through grant number 226375 
(ice2sea contribution number 022).
\section*{Appendix A. The two  fractional derivatives in $\RR^+$}
For a sufficiently well-behaved function $f(t)$
($ t\in \RR^+$) we may define the fractional derivative
of order $\beta  $ ($m-1 <\beta \le m\,,$ $\, m\in \NN$),
 see \eg the lectures notes by Gorenflo and Mainardi \cite{Gorenflo-Mainardi_CISM97} and 
 the text by Podlubny \cite{Podlubny_BOOK99}
in two different senses,  that we refer here as to
{\it Riemann-Liouville} derivative
and {\it Caputo} derivative, respectively.
Both derivatives are related to the so-called Riemann-Liouville
fractional integral of order $\alpha >0$,
  defined as
$$  I^\alpha  \, f(t) :=    \rec{\Gamma(\alpha )}\,
\int_0^t (t-\tau)^{\alpha-1} \, f(\tau )\, d\tau\,, \q \alpha >0\,.
\eqno(A.1)  $$
We note the convention $I^0 = Id$ (Identity operator)
and the semi-group property
$$ I^\alpha \, I^\beta = 
   I^\beta  \, I^\alpha = I^{\alpha +\beta} \,, \q
 \alpha\ge 0,\; \beta \ge 0\,. \eqno(A.2)$$
The fractional derivative of order $\beta >0$ in the 
{\it Riemann-Liouville} sense  is defined as the operator
$D^\beta$ which is the
left inverse of
the Riemann--Liouville integral of order $\beta $
(in analogy with the ordinary derivative), that is
$$ D^\beta \, I^\beta  = Id\,, \q \beta >0\,. \eqno(A.3) $$
If $m$ denotes the positive integer
such that $m-1 <\beta  \le m\,,$  we recognize from Eqs. (A.2) and (A.3)
$D^\beta  \,f(t) :=  D^m\, I^{m-\beta}  \,f(t)\,, $
hence
$$
 D^\beta  \,f(t) = 
 \, \left\{
 \begin{array}{ll}
  {\ds \frac{d^m}{dt^m}}\,
  {\ds \left[
   \rec{\Gamma(m-\beta )}\,\int _0^t
    \frac{f(\tau)}{(t-\tau )^{\beta  +1-m}}\, d\tau \right]	} ,
  &  \; m-1 <\beta  < m,\\
   {\ds \frac{d^m}{dt^m}\, f(t)} \,,
& \; \beta =m.
\end{array} \right .
\eqno(A.4)$$
For completion we define $D^0 = Id.$
\vsp
On the other hand, the fractional derivative of order $\beta >0$ in the
{\it Caputo} sense  is defined as the operator
$D_*^\beta$  such that
$    D_*^\beta \,f(t) :=  I^{m-\beta } \, D^m \,f(t)\,,$
hence
$$
    D_*^\beta \,f(t) = 
 \, \left\{ \begin{array}{ll}
    {\ds \rec{\Gamma(m-\beta)} }\, {\ds \int_0^t}
 {\ds \frac{f^{(m)}(\tau)} {(t-\tau )^{\beta+1-m}}\, d\tau}  \,,
&  \; m-1<\beta  <m,\\
   {\ds \frac{d^m}{dt^m} f(t)} \,, & \; \beta =m\,
\end{array} \right .
\eqno(A.5) $$
where $f^{(m)}$ denotes the derivative of order $m$ of the function $f$.
We point out the major utility
of the Caputo fractional derivative
in treating initial-value problems for physical and engineering
applications where initial conditions are usually expressed in terms of
integer-order derivatives. This can be easily seen
using the Laplace transformation, according to which
$$ \L \left\{ D_*^\beta \,f(t) ;s\right\} =
      s^\beta \,  \widetilde f(s)
   -\sum_{k=0}^{m-1}    s^{\beta  -1-k}\, f^{(k)}(0^+) \,,
  \q m-1<\beta  \le m \,,\eqno(A.6) $$
where
$ \widetilde f(s) =
\L \left\{ f(t);s\right\}
 = {\ds \int_0^{\infty}} \e^{\ds \, -st}\, f(t)\, dt\,, \;
s \in \CC$ and  $ f^{(k)}(0^+) := {\ds \lim_{t\to 0^+}}\, f^{(k)}(t)$.
The corresponding rule for the Riemann-Liouville
derivative is more cumbersome:  for $m-1<\beta  \le m $ it reads
$$ \L \left\{D^\beta  \, f(t);s\right\} =
      s^\beta \,  \widetilde f(s)
   -\sum_{k=0}^{m-1}\,
\left[D^k\, I^{(m-\beta )}\right]\,f(0^+) \, s^{m -1-k}\,,
\eqno(A.7)$$
where, in analogy with (A.6),  the limit for $t \to 0^+$
is understood to be taken after the operations of fractional integration
and derivation.  As soon as all the limiting   values $f^{(k)}(0^+)$
are finite
and $m-1 <\beta< m$, 
the formula (A.7)  simplifies into
$$ \L \left\{ D^\beta  \, f(t);s\right\} =
      s^\beta \,  \widetilde f(s) \,.\eqno(A.8)$$
In the special case   $f^{(k)}(0^+)=0$  for $k=0,1,  m-1$,
we recover the identity between the two fractional derivatives.
The Laplace transform rule (A.6)
was practically the starting point of Caputo, see 
\cite{Caputo_GJRAS67,Caputo_BOOK69},
in defining his generalized derivative in the late sixties of the last century.
\vsp
For further reading 
on the theory and applications of fractional calculus
we recommend the more recent treatise
by  Kilbas, Srivastava and Trujillo \cite{Kilbas-Srivastava-Trujillo_BOOK06}.
For the basic results on fractional integrability and differentiability 
we refer the interested reader to the survey paper by Li and Zhao \cite{li-zhao}.

\section*{Appendix B: Remark on initial conditions in classical and fractional operator equations}
   We note that the initial conditions at $t=0^+$,
$\sigma^{(h)}(0^+)$ with $h=0,1,\dots p-1$ and $\epsilon^{(k)}(0^+)$
with $k=0,1,\dots q-1$, do not appear in the operator equation but
they are required to be compatible  with the integral equations
(2.1). In fact, since Eqs. (2.1) do not contain the initial
conditions,
 some  compatibility conditions at $t=0^+$
must be   {\it implicitly} required both for stress and strain. In
other words, the equivalence between the integral Eqs. (2.1) and the
differential operator Eq. (2.14) implies that when we apply the
Laplace transform to both sides of Eq. (2.14) the contributions from
the initial conditions are vanishing or  cancel in pair-balance.
This can be easily checked for the simplest classical models
described by Eqs. (2.6)--(2.9). It turns out that the Laplace
transform of the corresponding constitutive equations does not
contain  any  initial conditions: they are all hidden being zero or
balanced between the RHS and LHS of the transformed equation. As
simple examples let us consider the Kelvin--Voigt model for which $p=0$,
$q=1$ and $m>0$, see Eq. (2.6), and the Maxwell model for which
$p=q=1$ and $m=0$, see Eq. (2.7).
\vsp
For the Kelvin--Voigt model we get 
$s\widetilde\sigma(s)=
 m\widetilde\epsilon(s) + b \left]s\widetilde\epsilon(s) -\epsilon(0^+)\right]$,
so,  for any causal stress and strain histories,
it would be
 $$ s\widetilde J(s) = \frac{1}{m +bs}   \iff \epsilon(0^+)=0\,.\eqno(B.1)$$
We note that the condition $\epsilon(0^+)=0$ is surely satisfied
for any reasonable stress history since $J_g=0$, but
is not valid for any reasonable strain history;
in fact, if we consider the relaxation test for which
$\epsilon(t)=\Theta(t)$ we have $\epsilon(0^+)=1$.
This fact may be  understood recalling that for the Voigt model
we have $J_g=0$ and $G_g=\infty$ (due to the delta contribution in the relaxation modulus).
\vsp
For the Maxwell model we get  
$\widetilde\sigma(s)+ a  \left[s\widetilde\sigma(s) -\sigma(0^+)\right]=
 b\left[s\widetilde\epsilon(s) -\epsilon(0^+)\right]$,
so,   for any causal stress and strain histories it would be
 $$ s\widetilde J(s) = \frac{a}{b}+ \frac{1}{bs} \iff a\sigma(0^+)= b \epsilon(0^+)\,. \eqno(B.2)$$
We now note that the condition $a\sigma(0^+)= b \epsilon(0^+)$ is
surely satisfied for any causal history both in stress and in strain.
This fact may be  understood recalling that for the Maxwell model
we have $ J_g >0$ and  $G_g = 1/J_g >0$.
\vsp
Then we can  generalize the above considerations stating that the compatibility
relations of the initial conditions are valid for all the four types of viscoelasticity,
as far as the creep representation is considered.
When the relaxation representation  is considered,
caution is required for the types III and IV, for which, for correctness, we would use the
generalized theory of integral transforms suitable just
for dealing with generalized functions.
\vsp
Similarly with the operator equations of integer order we note that the initial conditions at $t=0^+$
for the stress and strain
do not explicitly enter into the fractional operator equation (3.1).
This means
that the approach with  the Caputo derivative, which requires
in the Laplace domain the same initial conditions as the classical
models is quite correct. However, if we assume the same initial conditions,
the approach with the Riemann--Liouville derivative  provides
the same results since, in view of the corresponding Laplace transform rule (A.8),
 the initial conditions do not appear in the Laplace domain.
 The equivalence of the two approaches  was formerly outlined by Mainardi \cite{Mainardi_FDA04}
 and more recently noted by Bagley \cite{Bagley_FCAA07} for the fractional Zener model.
 \vsp
 We refer the reader to the paper by Heymans and Podubny \cite{Heymans-Podlubny_06}
   for the physical interpretation of initial conditions
   for fractional differential equations with Riemann-Liouville derivatives,
   especially in viscoelasticity.
   In such field, however, we prefer to adopt the Caputo derivative
   since it requires the same initial conditions
   as in the classical cases.
   By the way,  for  a physical view point,
   these initial conditions  are more   accessible than those required
   in the more general Riemann-Liouville approach, see (A.7).

\end{document}